\documentclass[conference]{IEEEtran}
\IEEEoverridecommandlockouts
\usepackage{amsmath}
\usepackage{amsfonts}
\usepackage{bbding}
\usepackage{amssymb}
\usepackage{array}
\usepackage{subfigure}

\usepackage{graphicx}
\usepackage{subfigure}
\usepackage[named]{algo}
\usepackage{algorithmic}
\usepackage{psfrag}
\usepackage{stfloats}
\usepackage[compress]{cite}
\makeatletter
\renewcommand{\citepunct}{,\penalty\@m\hskip.13emplus.1emminus.1em}
\renewcommand{\citedash}{\hbox{--}\penalty\@m}
\makeatother
\usepackage{setspace}
\usepackage{color}
\allowdisplaybreaks

\usepackage{amsthm}
\usepackage{stfloats}

\usepackage{hyperref}
\usepackage{bm}
\renewcommand{\IEEEQED}{\IEEEQEDopen}

\begin{document}
\title{Cross-layer Optimization for Ultra-reliable and Low-latency Radio Access Networks}

\author{
\IEEEauthorblockN{{Changyang She, Chenyang Yang and Tony Q.S. Quek}}
\thanks{Manuscript received June 25, 2016; revised January 06, 2017, April 12, 2017 and
	July 21, 2017; accepted Oct 2, 2017. The associate editor coordinating the review of this paper
	and approving it for publication was Q. Li.}
\thanks{This paper was presented in part at the 2016
	IEEE Global Communications Conference \cite{CYGC16}.}
\thanks{C. She was with the School of Electronics and
Information Engineering, Beihang University, Beijing 100191, China. He is now with the Information Systems Technology and Design Pillar, Singapore University of Technology and Design, 8 Somapah Road, Singapore 487372 (email:shechangyang@gmail.com).}
\thanks{C. Yang is with the School of Electronics and
	Information Engineering, Beihang University, Beijing 100191, China (email:cyyang@buaa.edu.cn).}
\thanks{C. She and C. Yang's work was supported in part by National Natural Science Foundation of China (NSFC) under Grant 61671036.}
\thanks{T. Q. S. Quek is with the Information Systems Technology and Design Pillar, Singapore University of Technology and Design, 8 Somapah Road, Singapore 487372 (e-mail: tonyquek@sutd.edu.sg).}
\thanks{C. She and T. Q. S. Quek's work was supported in part by was supported in part by the MOE ARF Tier 2 under Grant
	MOE2015-T2-2-104 and the SUTD-ZJU Research Collaboration under Grant SUTD-ZJU/RES/01/2016.}
\vspace{0.0cm}}

\maketitle \vspace{-0.0mm}
\begin{abstract}
In this paper, we propose a framework for cross-layer optimization to ensure ultra-high reliability and ultra-low latency in radio access networks, where both transmission delay and queueing delay are considered. With short transmission time, the blocklength of channel codes is finite, and the Shannon Capacity can not be used to characterize the maximal achievable rate with given transmission error probability. With randomly arrived packets, some packets may violate the queueing delay. Moreover, since the queueing delay is shorter than the channel coherence time in typical scenarios, the required transmit power to guarantee the queueing delay and transmission error probability will become unbounded even with spatial diversity. To ensure the required quality-of-service (QoS) with finite transmit power, a proactive packet dropping mechanism is introduced. Then, the overall packet loss probability includes \emph{transmission error probability}, \emph{queueing delay violation probability}, and \emph{packet dropping probability}. We optimize the packet dropping policy, power allocation policy, and bandwidth allocation policy to minimize the transmit power under the QoS constraint. The optimal solution is obtained, which depends on both channel and queue state information. Simulation and numerical results validate our analysis, and show that setting the three packet loss probabilities as equal causes marginal power loss.
\end{abstract}\vspace{-0.0 mm}

\vspace{-0.0mm}

\vspace{-3mm}\section{Introduction}
Supporting ultra-reliable and low-latency communications
(URLLC) has become one of the major goals in the fifth generation (5G) cellular networks \cite{3GPP2016Scenarios}. Ensuring such a stringent quality-of-service (QoS) enables various applications such as control of exoskeletons for patients, remote driving, free-viewpoint video, and synchronization of suppliers in a smart grid in tactile internet \cite{Gerhard2014The}, and autonomous vehicles and factory automation in ultra-reliable machine-type-communications (MTC) \cite{Popovski2014METIS}, despite that not all applications of tactile internet and MTC require both ultra-high reliability
and ultra-low latency.

Since tactile internet and MTC
are primarily applied for mission critical applications, the message  such as ``touch'' and control information is
usually conveyed in short packets, and the reliability is reflected by packet loss
probability \cite{3GPP2016Scenarios}. The traffic supported by URLLC distinguishes from traditional real-time service in both QoS requirement and packet size. For human-oriented applications, the requirements on delay and reliability are medium. For example, in the long term evolution (LTE) systems, the maximal queueing delay and its violation probability for VoIP are respectively $50$~ms and $2\times 10^{-2}$ in radio access networks, and the minimal packet size is 1500 bytes \cite{3GPPQoS}. For control-oriented applications such as vehicle collision avoidance or factory automation, the end-to-end (E2E) or round-trip delay is around 1 ms, the overall packet loss probability is $10^{-5}\sim 10^{-9}$  \cite{Gerhard2014The,A2014Scenarios}, and the packet size is 20 bytes or even smaller \cite{3GPP2016Scenarios}.

LTE systems were designed for human-oriented applications, where the E2E delay includes uplink (UL) and downlink (DL) transmission delay, coding and processing delay, queueing delay, and routing delay in backhaul and core networks \cite{Shao2015ultra}. The radio resources are allocated in every transmit time interval (TTI), which is set to be $1$~ms \cite{Capozzi2013Downlink}. This means that the packets need to wait in the buffer of base station (BS) more than $1$~ms before transmission. Therefore, even if other delay components in backhaul and core networks are reduced with new network architectures \cite{Meryem2016Tactile},  LTE systems cannot ensure the E2E or round-trip latency of 1 ms.

\subsection{Related Work}
While reducing latency in wireless networks is challenging, further ensuring high reliability
makes the problem more intricate. To reduce the delay  caused by transmission and signalling \cite{Shehzad2015Control}, a short frame structure was introduced in \cite{Petteri2015A}, and the TTI was set identical to the frame duration.
To ensure high reliability of transmission with short frame, proper channel coding with finite blocklength is important. Fortunately, the results in \cite{Yury2010Channel} indicate that it is possible to guarantee very low transmission error probability with short blocklength channel codes, at the expense of achievable rate reduction. By using practical coding schemes like Polar codes \cite{Kai2014Polar}, the delays caused by transmission, signal processing and coding can be reduced.

Exploiting diversity among multiple links has long been used as an effective way to improve the successful transmission probability in wireless communications. To support the high reliability over fading channels, various diversity techniques have been investigated, say frequency diversity and macroscopic diversity in single antenna systems  \cite{David2014Achieving,Felix2015diversity} and  spatial diversity in multi-antenna systems \cite{Beatriz2015Reliable}. Simulation results using practical modulation and coding schemes in \cite{Osman2015Analysis,Niklas2015Radio} show that the required transmit power to ensure given transmission delay and reliability can be rapidly reduced when the number of antennas at a BS increases.

In all these works, only transmission delay and transmission error probability are taken into account in the QoS requirement. In practice, since the packets arrive at the buffer of the BS randomly, there is a queue at the BS. To control the delay and packet loss caused by both queueing and transmission, cross-layer optimization should be considered \cite{CYGC16}. Similar to the real-time service such as VoIP, the required queueing performance of
URLLC can be modeled as statistical queueing requirement, characterized by the maximal
queueing delay and a small delay violation probability. By using effective bandwidth  \cite{EB} and effective capacity  \cite{EC} to analyze performance of tactile internet under the  statistical queueing requirement, the tradeoff among queueing delay, queueing delay violation probability and throughput was studied in \cite{Beatriz2014Tradeoffs}, and UL and DL resource allocation was jointly optimized to achieve the E2E delay requirement in \cite{Adnan2016Towards}. In both works, the Shannon capacity is applied to derive the effective capacity. However, with short transmission delay requirement, channel coding is performed with a finite block of symbols, with which the Shannon capacity is not achievable. In fact, the results obtained  by using network calculus  in \cite{Gross2015Delay} show that if Shannon capacity is used to approximate the achievable rate of
short blocklength codes for designing resource allocation, the queueing delay and delay violation probability cannot be guaranteed.

Based on the achievable rate of a single antenna system with finite blocklength channel codes derived in \cite{Yury2010Channel}, queueing delay/length was analyzed in \cite{Throughput2011Mustafa,Shengfeng2015Convexity}. For applications with medium delay and reliability requirements, the throughput subject to statistical queueing constraints was studied in \cite{Throughput2011Mustafa}, where the effective capacity was derived by using the achievable rate with finite blocklength channel codes, and an automatic repeat-request (ARQ) mechanism was employed to improve reliability.
 An energy-efficient packet scheduling policy was optimized  in \cite{Shengfeng2015Convexity} to ensure a strict deadline by assuming packet arrival time and instantaneous channel gains known \emph{a prior}, while the deadline violation probability under the transmit power constraint was not studied.

\subsection{Major Challenges and Our Contributions}
Supporting URLLC leads to the following challenges in radio resource allocation.

First, the required queueing delay and transmission delay are shorter than channel coherence time in typical scenarios of URLLC.\footnote{In this scenario, effective capacity can no longer be applied.} This results in the following problems. (1) ARQ mechanism can no longer be used to improve reliability. This is because retransmitting a packet in subsequent frames not only introduces extra transmission delay but also can hardly improve the successful transmission probability when the channels in multiple frames stay in deep fading. (2) Time diversity cannot be exploited to enhance reliability, and frequency diversity may not be scalable to the large number of nodes. Moreover, whether spatial diversity can guarantee the reliability is unknown. (3) The studies in \cite{Berry2013} show that when the average delay approaches the channel coherence time, the average transmit power could become infinity, because transmitting packets during deep fading leads to unbounded transmit
power. Hence, how to ensure both the ultra-low delay and the ultra-high reliability with finite
transmit power is unclear.

Second, the blocklength of channel codes is finite. The maximal achievable rate in finite blocklength regime is neither convex nor concave in radio resources such as transmit power and bandwidth \cite{Yury2010Channel,Yury2014Quasi}. As a result, finding optimal resource allocation policy for URLLC is much more challenging than that for traditional communications, where Shannon capacity is a good approximation of achievable rate and is jointly concave in transmit power and bandwidth.

Third, effective bandwidth is a powerful tool for designing resource allocation to satisfy the statistical queueing requirement of real-time service \cite{EB}.
Since the distribution of queueing delay is obtained based on large deviation principle, the effective bandwidth can be used when the delay bound is large and the delay violation probability is small \cite{Ward1993Tail}. Therefore, using effective bandwidth for URLLC seems problematic.

In this paper, we propose a cross-layer optimization framework for URLLC. While technical challenges in achieving ultra-low E2E/round-trip delay exist at various levels, we only consider transmission delay and queueing delay in radio access networks, and focus on DL transmission. The major contributions of this work are summarized as follows:
\begin{itemize}
\item We show that only exploiting spatial diversity cannot ensure the ultra-low latency and ultra-high reliability with finite transmit power  over fading channels. To ensure the QoS with finite transmit power, we propose a proactive packet dropping mechanism.
\item We establish a framework for cross-layer optimization to guarantee the low delay and high reliability, which includes a resource allocation policy and the proactive packet dropping policy depending on both channel and queue state information. By assuming frequency-flat fading channel model, we first optimize the power allocation and packet dropping policies in a single-user scenario, and then extend to the multi-user scenario by further optimizing bandwidth allocation among users. Moreover, how to apply the framework to frequency-selective channel is also discussed.
\item We validate that even when the delay bound is extremely short, the upper bound of the complementary cumulative distributed function (CCDF) of queueing delay derived from effective bandwidth still works for Poisson process and Interrupted Poisson Process (IPP), which is more bursty than Poisson process, and Switched Poisson Process (SPP), which is an autocorrelated two-phase Markov Modulated Poisson Process \cite{Jian2015IPP}.
\item We consider the \emph{transmission error probability} with finite blocklength channel coding, the \emph{queueing delay violation probability}, and the \emph{proactive packet dropping probability} in the overall reliability. By simulation and numerical results, we show that setting packet loss probabilities equal is a near optimal solution in terms of minimizing transmit power.
\end{itemize}

The rest of this paper is organized as follows. Section II describes system model and QoS requirement. Section III shows how to represent queueing delay constraint with effective bandwidth. Section IV introduces the packet dropping policy, and the framework for cross-layer optimization. Section V illustrates how to apply the framework to frequency-selective channel. Simulation and numerical results are provided in Section VI to validate our analysis and to show the optimal solution. Section VII concludes the paper.

\section{System Model and QoS Requirement}
Consider a frequency division duplex cellular system,\footnote{Our studies can be easily extended into time division duplex system, which is with different short frame structure \cite{Petteri2015A}.} where each BS with $N_\mathrm{t}$ antennas serves $K+M$ single-antenna nodes. The nodes are divided into two types. The first type of nodes are $K$ users, which need to upload packets and download packets from the BS. The second type of nodes are $M$ sensors, which only upload packets. In the cases without the need to distinguish between users and sensors, we refer both as nodes. Time is discretized into frames. Each frame consists of a data transmission phase and a phase to transmit control signaling (e.g., pilot for channel estimation). We consider frequency reuse among adjacent cells and orthogonal frequency division multiple access (OFDMA) to avoid interference.

\begin{figure}[htbp]
	\vspace{-0.4cm}
	\centering
	\begin{minipage}[t]{0.45\textwidth}
		\includegraphics[width=1\textwidth]{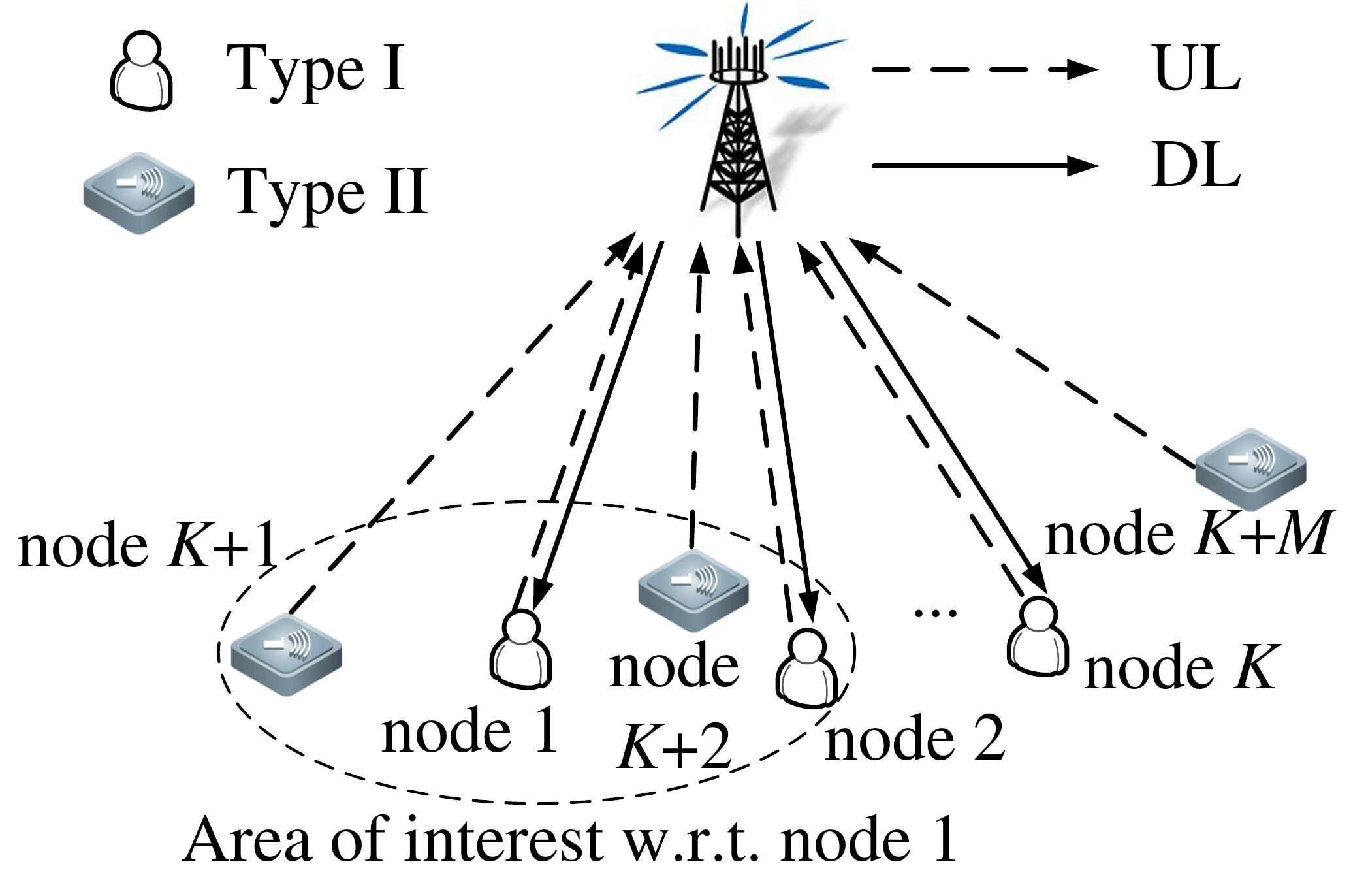}
	\end{minipage}
	\vspace{-0.2cm}
	\caption{System model.}
	\label{fig:smodel}
	\vspace{-0.4cm}
\end{figure}

All nodes in a cell upload their messages with short packets to the BS. The BS processes the received messages from the nodes, and then transmits the relevant messages to the target users. For example, nodes $2$, $K+1$, and $K+2$ lie in the area of interest with respect to (w.r.t.) user $1$, as shown in Fig. \ref{fig:smodel}, and the BS only transmits the messages from nodes $2$, $K+1$, and $K+2$ to user $1$. Such system model can be applied in analyzing E2E delay in local communication scenarios, where all nodes are associated to adjacent BSs that are connected with each other by fiber backhaul. The delay in fiber backhaul is much less than $1$~ms \cite{Tony2016Backhaul}, and hence the delay in radio access network dominates the E2E delay. For other communication scenarios (e.g., remote control), the delay components in backhaul and core networks should be taken into account, yet our model can still be used to analyze the delay in radio access \cite{3GPP2016Scenarios}. Moreover, the model captures one of the key features of ultra-reliable MTC \cite{Popovski2014METIS}: a packet generated by one node may be required by multiple users, and one user may also require packets generated by multiple nodes. Hence the model is representative for URLLC, although it cannot cover all application scenarios.\footnote{Direct transmission between nodes (i.e., device-to-device (D2D) communication mode) can help reduce delay with only one hop transmission. However, in D2D mode, the interference becomes more complex than the centralized communications \cite{Daquan2014Device}. How to use D2D mode for URLLC  deserves further study but is beyond the scope of this work.} All the notations to be used throughout the paper are summarized in Table I.

\begin{table*}[htbp]
\scriptsize
\renewcommand{\arraystretch}{1.3}
\caption{Summary of notations}
\begin{center}\vspace{-0.2cm}
\begin{tabular}{|p{0.8cm}|p{7cm}||p{0.8cm}|p{7cm}|}
\hline
  $K$ & number of users  &
  $M$ & number of sensors  \\\hline
   $T_{\rm c}$ & channel coherence time& $T_{\rm f}$ & duration of one frame \\\hline
  $D_{\max}$ & required delay bound in radio access network & $D^q_{\max}$ & queueing delay bound\\\hline
  $\phi$ & duration for data transmission in each frame&  $N_{\rm t}$ & number of antennas at the BS
  \\\hline
  $\varepsilon_k^q$ & queueing delay violation probability of the $k$th user &
  $\varepsilon_c^q$ & transmission error probability of the $k$th user\\\hline
  $\varepsilon_c^h$ & proactive packet dropping probability of the $k$th user &
  ${\varepsilon _\mathrm{D}}$ & overall packet loss probability\\\hline
   $N_k^{\rm sc}$ & number of subchannels allocated to the $k$th user &
  $N^{\rm c}_k$ & number of subcarriers allocated to the $k$th user\\\hline
  $W_{\rm c}$ & bandwidth of each subchannel & $B$ & bandwidth of each subcarrier \\\hline
  $n_k^s$ & blocklength of channel coding of the $k$th user &$W_k$ &total bandwidth allocated to the $k$th user
   \\\hline
  $s_k(n)$ & achievable rate with finite blocklength of the $k$th user in the $n$th frame &
  $s_k^\infty(n)$ & capacity of the $k$th user in the $n$th frame\\\hline
  ${\bf{h}}_k$ & channel vector of the $k$th user &
  $\mu_k$ & average channel gain of the $k$th user\\\hline
  $g_k$ & normalized instantaneous channel power gain of the $k$th user&$P_k(n)$ & transmit power allocated to the $k$th user in the $n$th frame\\\hline
  $N_0$ & single-sided noise spectral density &
  $u$ & number of bits in one packet \\\hline
  $f_{\rm Q}^{-1}(x)$ & inverse of Q-function & $f_g(x)$ & probability density function of normalized instantaneous channel gain \\\hline
  ${\mathcal{A}}_k$ & a set consists of the indices of the nodes that lie in the area of interest w.r.t. the $k$th user &
  $a_i(n)$ &  the number of packets uploaded to the BS from the $i$th node \\\hline
  $b_k(n)$ & number of packets departed from the $k$th queue in the $n$th frame &
  $Q_k\left( n \right)$ & queue length of the $k$th user in the $n$th frame \\\hline
  $E^B_k(\theta_k)$ & effective bandwidth of the arrival process to the $k$th user & $\theta_k$ & the QoS exponent of the $k$th user \\\hline
  $P^{\rm UB}_{D_k}$ & upper bound of queueing delay violation probability of the $k$th queue & $\pi_l$ & probability that there are $l$ packets in the queue \\\hline
  $\lambda_k$ & average packet rate of the $k$th Poisson process& $\lambda^{\rm on}_k$ & average packet rate in the ``ON" state of the $k$th IPP \\\hline
  $\alpha^{-1}$ & average duration of ``OFF" state of IPP& $\beta^{-1}$ & average duration of ``ON" state of IPP \\\hline
  $\alpha_{\rm I}^{-1}$ & average duration of the first state of SPP& $\alpha_{\rm II}^{-1}$ & average duration of the second state of SPP \\\hline
  $\lambda_k^{\rm I}$ & average packet rate in the first state of the $k$th SPP& $\lambda^{\rm II}_k$ & average packet rate in the second state of the $k$th SPP \\\hline
  $\xi_k$ & ratio of average arrival rate to service rate of the $k$th queue & $\gamma_k$ & required SNR of the $k$th user \\\hline
   $\eta_k$ & buffer non-empty probability of the $k$th queue& $P^{\rm th}_k$ & maximal transmit power that can be allocated to the $k$th user \\\hline
\end{tabular}
\end{center}
\vspace{-0.2cm}
\end{table*}

\subsection{QoS Requirement}
The QoS requirement of  each
user is characterized by the E2E delay and overall  loss probability for each packet \cite{3GPP2016Scenarios,Popovski2014METIS}. In the considered radio access network, the E2E delay bound, denoted as $D_{\max}$, includes UL and DL transmission delay and queueing delay. We only consider one-way delay requirement. By setting $D_{\max}$ less than half of round-trip delay, our study can be directly extended to the applications with requirement on round-trip delay.

To ensure ultra-low transmission delay, we consider the short frame structure proposed  in \cite{Shehzad2015Control}, where the TTI is equal to the frame duration $T_\mathrm{f}$, each consisting of a duration for data transmission $\phi$ and a duration for control signalling, as shown in Fig. \ref{Illustration}.  Owing to the required short delay, $T_\mathrm{f} \ll D_{\max}$, and retransmission mechanism is unable to be used. Both UL transmission and DL transmission of each short packet are finished within one frame, respectively. If a packet is not transmitted error-free in one frame, then the packet will be lost. Because only a few symbols can be transmitted within $\phi$, the transmission error is not zero with finite blocklength channel codes among these symbols. Since UL transmission has been studied in \cite{She2016GCworkshop}, we focus on the DL transmission in this work. Then, the overall reliability for each
user, denoted as ${\varepsilon _\mathrm{D}}$, is the overall packet loss
probability minus the UL transmission error probability. Denote the DL transmission error probability (i.e. the block error probability \cite{Yury2014Quasi}) for the $k$th user as ${\varepsilon_k^c}$.

Since the UL and DL transmissions need two frames, the queueing delay for every packet should be bounded as $D^q_{\max} \triangleq D_{\max}-2T_\mathrm{f}$. If the queueing delay bound is not satisfied, then a packet will become useless and has to be dropped. Denote the reactive packet dropping probability due to queueing delay violation as ${\varepsilon_k^q}$. As detailed later, to satisfy the requirement imposed on the queueing delay for each packet  $(D^q_{\max}, {\varepsilon_k^q})$ and ${\varepsilon_k^c}$ to the $k$th user, the required transmit power may become unbounded in deep fading. To guarantee QoS with finite transmit power, we proactively drop several packets in the queue under deep fading and control the overall reliability. Denote the proactive packet dropping probability  for the $k$th user as  ${\varepsilon_k^h}$.

Then, the overall reliability for the $k$th user can be characterized by the overall packet loss probability, which is
\begin{align}
&1 - (1- {\varepsilon_k^c})(1- {\varepsilon_k^q})(1- {\varepsilon_k^h})\approx {\varepsilon_k^c} + {\varepsilon_k^q} +{\varepsilon_k^h} \leq {\varepsilon _\mathrm{D}},\label{eq:reliability}
\end{align}
where the approximation is accurate since ${\varepsilon_k^c}$, ${\varepsilon_k^q}$, and ${\varepsilon_k^h}$ are extremely small.


\subsection{Channel Model}
\begin{figure}[htbp]
        \vspace{-0.3cm}
        \centering
        \begin{minipage}[t]{0.5\textwidth}
        \includegraphics[width=1\textwidth]{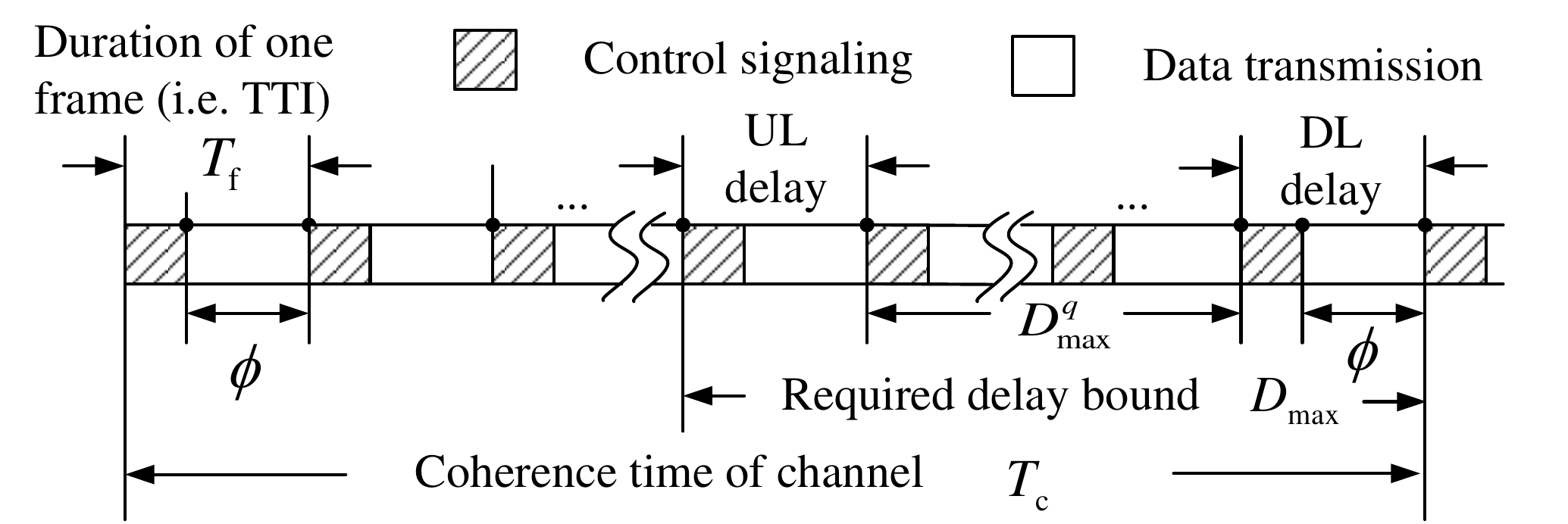}
        \end{minipage}
        \vspace{-0.2cm}
        \caption{Relation of the required delay bound, channel coherence time, frame duration and TTI. The UL transmission delay is equal to $T_{\rm f}$, and the same to the DL transmission delay.}
        \label{Illustration}
        \vspace{-0.3cm}
\end{figure}

We consider block fading, where the channel remains constant within a coherence interval and varies independently among intervals. Denote the channel coherence time as $T_{\rm c}$. Since the required delay bound $D_{\max}$ is very short, it is reasonable to assume that $T_{\rm c} > D_{\max} > D^q_{\max}$, as shown in Fig. \ref{Illustration}.\footnote{For instance, for users with velocities less than $120$~km/h in a vehicle communication system operating in carrier frequency of $2$~GHz, the channel coherence time is larger than $1$~ms, which exceeds the delay bound of each packet. For other applications like smart factory, the velocities of sensors are slow or even zero, and hence $T_{\rm c} \gg 1$~ms.} In the following, we consider such a representative scenario for typical  applications of URLLC, which is more challenging than the other case with $T_{\rm c} \leq D^q_{\max}$. Since $T_{\rm f}$ should be less than $D_{\max}$ and the channel coding is performed within $\phi$ of each frame, such a channel (i.e., $T_{\rm f} <T_{\rm c}$) is referred to as \emph{quasi-static fading channel} as in \cite{Yury2014Quasi}.

Denote the average channel gain of the $k$th user as $\mu_k$, and the corresponding channel vector in a certain coherence interval as ${\bf{h}}_k \sim {\cal CN}(0,1) \in {\mathbb{C}}^{N_{\rm t} \times 1}$ with independent and identically distributed (i.i.d.) zero mean and unit variance Gaussian elements. Denote the size of each packet as $u$~bits. According to the Shannon capacity formula with infinite blocklength coding, when $\mu_k$ and  ${\bf{h}}_k$ are perfectly known at the BS, the maximal number of packets that \emph{can be} transmitted to the $k$th user in the $n$th frame can be expressed as
\begin{align}
s^\infty_k(n) = \frac{ \phi B N^{\rm c}_k }{u \ln{2}} \ln\left[1+\frac{\mu_k P _k(n)g_k}{ N_0 B N^{\rm c}_k}\right] \;
\text{(packets)}, \label{eq:Cn}
\end{align}
where  $P _k(n)$ is the transmit power allocated to the $k$th user in the $n$th frame, $g_k = {\bf{h}}_k^H{\bf{h}}_k$,  $N_0$ is the single-sided noise spectral density, $B$ is the separation among subcarriers, $N^{\rm c}_k$ is number of subcarriers allocated to the $k$th user, and $[ \cdot ]^H$ denotes the conjugate transpose. When the bandwidth allocated to the $k$th user, $W_k = B N^{\rm c}_k$, is smaller than coherence bandwidth, the channel is flat fading and the channel gains over $N^{\rm c}_k$ subcarriers are approximately identical. We first consider flat fading channel, which is applicable for many scenarios of tactile internet and utra-reliable MTC  where the number of users is large. We then discuss how to apply the proposed framework to frequency-selective channels in Section V.

The number of symbols transmitted in one frame (also referred to as the blocklength of channel coding)  for the $k$th user,  $n^s_k$, is determined by the bandwidth and duration, i.e. $n^s_k = \phi W_k$. To ensure the ultra-low latency, the transmission duration $\phi$ is very short. Considering that the bandwidth for each user is limited, $n^s_k$ is far from infinite, and hence $s^\infty_k(n)$ is not achievable. The maximal achievable rate with finite blocklength coding is with very complicated expression \cite{Yury2014Quasi}. By using the \emph{normal approximation} in \cite{Yury2014Quasi}, the maximal number of packets that \emph{can be} transmitted to the $k$th user in the $n$th frame can be accurately approximated as
\begin{align}
s_k(n) \approx \frac{ \phi B N^{\rm c}_k}{u \ln{2}} &\left\{\ln\left[1+\frac{\mu_k P _k(n)g_k}{ N_0 B N^{\rm c}_k}\right] - \sqrt{\frac{V_k}{\phi B N^{\rm c}_k}}f_{\rm Q}^{-1}({\varepsilon_k^c})\right\}\;\nonumber\\
&\quad\quad\quad\quad\quad\quad\quad\quad\quad\quad\quad\quad\text{(packets)}, \label{eq:sn}
\end{align}
where  $f_{\rm Q}^{-1}(x)$ is the inverse of Q-function, and $V_k$ is given by \cite{Yury2014Quasi}
\begin{align}\label{eq:dispersion}
V_k = 1-\frac{1}{\left[1+\frac{\mu_k P _k(n)g_k}{ N_0 B N^{\rm c}_k}\right]^2}.
\end{align}
\eqref{eq:sn} is obtained for interference-free systems, which is valid for the considered OFDMA (and also for time division multiple access or space division multiple access with zero-forcing beamforming). To consider other multiple access techniques where interference cannot be completely avoided, the achievable rate with finite blocklength in interference channels
should be used, which however is not available in the literature until now.

As shown in \cite{Gross2015Delay}, if \eqref{eq:Cn} is used to design resource allocation with finite blocklength coding, then the queueing delay and the queueing delay violation probability will be underestimated. As a result, the allocated resource is insufficient for ensuring the queueing performance. This indicates that to guarantee ultra-low latency and ultra-high reliability, \eqref{eq:sn} should be applied.

\subsection{Queueing Model}
\begin{figure}[htbp]
        \vspace{-0.2cm}
        \centering
        \begin{minipage}[t]{0.5\textwidth}
        \includegraphics[width=1\textwidth]{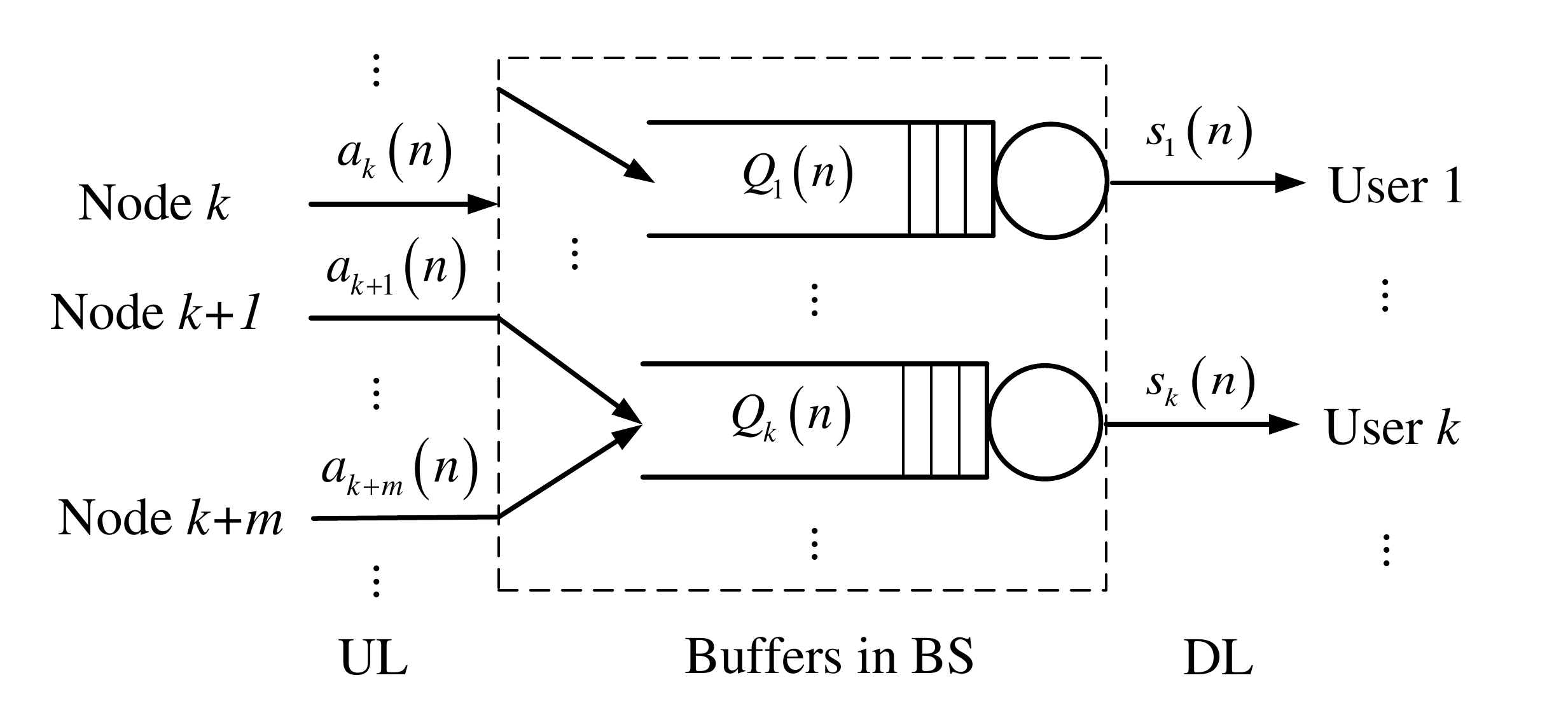}
        \end{minipage}
        \vspace{-0.3cm}
        \caption{Queueing model at the BS.}
        \label{fig:Qmodel}
        \vspace{-0.2cm}
\end{figure}

In the $n$th frame, the $k$th user requests the packets uploaded from its nearby nodes. The indices of the nodes that lie in the area of interest w.r.t. the $k$th user constitute a set ${\mathcal{A}}_k$ with cardinality $\left|{\mathcal{A}}_k\right|$. As illustrated in Fig. \ref{fig:Qmodel}, the index set of the nearby nodes of the $k$th user is ${\mathcal{A}}_k = \{k+1,...,k+m\}$.
Then, the number of packets waited in the queue for the $k$th user at the beginning of the $(n+1)$th frame can be expressed as
\begin{align}
Q_k\left( {n + 1} \right) = \max \left\{ {Q_k\left( n \right) - s_k\left( n \right)  },0 \right\}+ \sum\limits_{i \in {\mathcal{A}}_k} {{a_i}\left( n \right)}, \label{eq:queue}
\end{align}
where $a_i\left( n \right)$, $i \in {\mathcal{A}}_k$ is the number of packets uploaded to the BS from the $i$th nearby node of the $k$th user.

We  consider the scenario that the inter-arrival time between packets could be shorter than $D^q_{\max}$ (otherwise the queueing delay is zero), which happens when the packets for a target user are randomly uploaded from multiple nearby nodes, i.e. $\left|{\mathcal{A}}_k\right|>1$. At the first glance, such a scenario seems to occur with a low probability. However, to ensure the ultra-high reliability of ${\varepsilon _D}=0.001$\%$\sim$$0.00001$\%, the scenario of non-zero queueing delay is not negligible. Denote the number of packets departed from the $k$th queue in the $n$th frame as $b_k(n)$. If all the packets in the queue can be completely transmitted in the $n$th frame, then $b_k(n) = Q_k(n)$. Otherwise, $b_k(n)=s_k(n)$. Hence, we have
\begin{align}
b_k(n) = \min \left\{Q_k\left( n \right), s_k\left( n \right)\right\}. \label{eq:bn}
\end{align}

Using \eqref{eq:queue} and \eqref{eq:bn}, the evolution of the queue length can be described as follows,
\begin{align}
Q_k\left( {n + 1} \right) - Q_k\left( n \right) = \sum\limits_{i \in {\mathcal{A}}_k} {{a_i}\left( n \right)} - b_k(n). \label{eq:DeltaQ}
\end{align}

\section{Ensuring the Queueing Delay Requirement}
In this section we employ {effective bandwidth} to represent the queueing delay requirement. We validate that effective bandwidth can be applied in the short delay regime for Poisson arrival process, and then extend the discussion to IPP and SPP.

\subsection{Representing Queueing Delay Constraint with Effective Bandwidth} For stationary packets arrival process $\{{\sum\limits_{i \in {\mathcal{A}}_k}{{a_i}\left( n \right)} }, n = 1,2,...\}$, the effective bandwidth is defined as \cite{EB}
\begin{align}
E^{B}_k(\theta_k)
= \mathop {\lim }\limits_{N \to \infty } \frac{1}{{NT_{\rm f}{\theta _k}}}&\ln \left\{ {{\mathbb{E}}\left[ {\exp \left( {{\theta _k}\sum\limits_{n = 1}^N {\sum\limits_{i \in {{\cal A}_k}} {{a_i}\left( n \right)} } } \right)} \right]} \right\}\nonumber\\
&\quad\quad\quad\quad\quad\quad\quad\text{(packets/s)},\label{eq:EB}
\end{align}
where $\theta_k$ is the QoS exponent for the $k$th user. A larger value of $\theta_k$ indicates a smaller queueing delay bound with given queueing delay violation probability.

{\bf{Remark 1:}} When the queueing delay bound is not longer than the channel coherence time, the service process is constant within the delay bound with given resources such
as transmit power and bandwidth, and the power allocation over fading channel is channel inversion in order to guarantee queueing delay \cite{Tang2007Quality}. This is also true when achievable rate in \eqref{eq:sn} is applied, as explained in what follows. To satisfy the queueing delay requirement of the $k$th user $(D^q_{\max}, {\varepsilon_k^q})$ in fading channels, the constant service rate should be no less than the effective bandwidth of the arrival process of the user. By setting $s_k(n)$ in \eqref{eq:sn} equal to $E^{B}_k(\theta_k)$, $P_k(n)g_k$ is constant, i.e., the power allocation is channel inversion, which is not always feasible in practical fading channels. We will show how to handle this issue in the next section.

When the $k$th user is served with a constant rate equal to $E^{B}_k(\theta_k)$, the steady state queueing delay violation probability can be approximated as \cite{EC}
\begin{align}
\Pr\{D_k(\infty) > D^q_{\max}\} \approx \eta_k \exp\{-\theta_k E_k^B(\theta_k) D^q_{\max}\},\label{eq:apporxD}
\end{align}
where $\eta_k$ is the buffer non-empty probability and the approximation is accurate when $D^q_{\max} \to \infty$ (i.e. queue length is large enough)  \cite{EB}. Since $\eta_k \leq 1$, we have
\begin{align}
\Pr\{D_k(\infty) > D^q_{\max}\} \leq \exp\{-\theta_k E_k^B(\theta_k) D^q_{\max}\} \triangleq P_{D_k}^{\rm UB}. \label{eq:UB}
\end{align}
If the upper bound in \eqref{eq:UB} satisfies
\begin{align}
P_{D_k}^{\rm UB} = \exp\{-\theta_k E_k^B(\theta_k) D^q_{\max}\} = \varepsilon_k^q, \label{eq:delay}
\end{align}
then the queueing delay requirement $(D^q_{\max}, \varepsilon_k^q)$ can be satisfied.
In other words,
if the number of packets transmitted in every frame to the $k$th user is a constant that satisfies
\begin{align}
s_k(n) = T_{\rm f}E_k^B(\theta_k)\; (\text{packets}), \label{eq:QoS}
\end{align}
then $(D^q_{\max}, \varepsilon_k^q)$ can be ensured  \cite{EB}. When the $k$th queue is served by the constant service process $\{s_k(n), n = 1,2,...\}$ that satisfies \eqref{eq:QoS}, the departure process in  \eqref{eq:bn} becomes
\begin{align}
b_k(n) = \min\{Q_k(n),T_{\rm f}E^B_k(\theta_k)\}\; (\text{packets}). \label{eq:QoSbn}
\end{align}
If the departure process  $\{b_k(n), n = 1,2,...\}$ satisfies \eqref{eq:QoSbn}, then $(D^q_{\max}, \varepsilon_k^q)$ can be guaranteed. Satisfying \eqref{eq:QoSbn} does not require constant service process. For example, when $Q_k(n) = 0$, the buffer is empty, then no service is needed.

\subsection{Validating the Upper Bound $P_{D_k}^{\rm UB}$ in \eqref{eq:UB} with Representative Arrival Processes}
\subsubsection{Representative arrival processes} The aggregation of packets that are independently generated by $\left|{\mathcal{A}}_k\right|$ nodes lie in the concerned area w.r.t the $k$th user  (i.e. $\sum\limits_{i \in {\mathcal{A}}_k} {{a_i}\left( n \right)}$ in \eqref{eq:queue}) can be modeled as a Poisson process in vehicle communication and other MTC applications \cite{Mehdi2013Performance,3GPP2012MTC}. Denote the average packet rate of the $k$th Poisson process as $\lambda_k$.

Since the features of traffic, say burstiness and autocorrelation, have large impact on the
delay performance of queueing systems  \cite{Jian2015IPP,Gross1985MD1}, and the effective bandwidth for real-world arrival
processes is hard to obtain, we also consider another two representative traffic models.

As shown in \cite{Hassan2013A}, the event-driven packet arrivals in vehicular communication networks can be modelled as a bursty process, IPP. When no event happens, no sensor sends packets to the BS. When an event happens (e.g., a sudden brake) and detected by nearby sensors, the sensors send the packets to the BS.
IPP has two states. In the ``OFF" state, no packet arrives. In the ``ON" state, packets arrive at the buffer of the BS according to a Poisson process with average packet rate $\lambda^{\rm on}_k$~packets/frame. The durations that the process stays in ``OFF" and ``ON" states are exponential distributed with mean values of $\alpha^{-1}$ and $\beta^{-1}$ frames, respectively.

Both Poisson process and IPP are renewal processes, which cannot characterize the autocorrelation of a traffic. In \cite{Hassan2013A}, SPP is used to model the aggregation of event-driven packets and periodic packets in vehicle communication networks. Similar to IPP, SPP has two states, where the durations that a SPP stays in the first state and the second state are exponential distributed with mean values of $\alpha_{\rm I}^{-1}$ and $\alpha_{\rm II}^{-1}$ frames, respectively. In the two states, packets arrive at the buffer of the BS according to Poisson processes with average packet rates $\lambda^{\rm I}_k$ and $\lambda^{\rm II}_k$~packets/frame, respectively. Therefore, a SPP is determined by parameters $(\lambda^{\rm I}_k,\lambda^{\rm II}_k,\alpha_{\rm I},\alpha_{\rm II})$.

The effective bandwidths of Poisson process, IPP and SPP are provided in Appendix \ref{App:EB}.


\subsubsection{Validating the upper bound}
The approximation in \eqref{eq:apporxD} is accurate when the delay bound is sufficiently large and $\varepsilon_k^q$ is very small \cite{EB,Ward1993Tail}. However, it is unclear how large $D^q_{\max}$ needs to be for an accurate approximation. One possible reason is that it is very difficult to obtain an accurate distribution of the queueing delay.

In fact, what really concerned here is whether the upper bound in \eqref{eq:UB} is applicable to our problem. If $P_{D_k}^{\rm UB}$ is indeed an upper bound of $\Pr\{D_k(\infty) > D^q_{\max}\}$, then a transmit policy optimized under the constraint in \eqref{eq:QoS} or \eqref{eq:QoSbn} can satisfy the queueing delay requirement.
In what follows, we derive the queueing delay distribution for Poisson process, which can be used to validate the upper bound in short $D_{\max}^q$ regime numerically. For arrival processes that are more bursty than Poisson process, the upper bound in \eqref{eq:UB} is applicable \cite{squeezing1996}.

When a Poisson arrival process is served by a constant service process $\{s_k(n), n = 1,2,...\}$, the well-known M/D/1 queueing model can be applied \cite{Gross1985MD1}. For a discrete state M/D/1 queue, the CCDF of the steady state queue length can be expressed as $\Pr\{Q_k(\infty) > L\} = 1 - \sum\limits_{l = 1}^L {{\pi _l}}$,
where $\pi_l = \Pr\{Q_k(\infty) = l\}$ is the probability that there are $l$ packets in the queue, i.e.,
\begin{align}
&\pi_0 = 1-{\xi_k},\;\pi_1 = (1-{\xi_k})(e^{{\xi_k}}-1),\nonumber\\
&\pi_l = (1-{\xi_k}) \times \nonumber\\
&\left\{e^{l {\xi_k}} +\sum\limits_{j = 1}^{l - 1} {{e^{j{\xi_k} }}{{\left( { - 1} \right)}^{l - j}}\left[ {\frac{{{{\left( {j{\xi_k} } \right)}^{l - j}}}}{{\left( {l - j} \right)!}} + \frac{{{{\left( {j{\xi_k} } \right)}^{l - j - 1}}}}{{\left( {l - j - 1} \right)!}}} \right]} \right\}, \nonumber\\
&(l \geq 2),\label{eq:pdfMD1}
\end{align}
with ${\xi_k} = \lambda_k / s_k(n)$ \cite{Gross1985MD1}. For a Poisson arrival process served by a constant service rate $ \frac{1}{T_{\rm f}}s_k(n) = E^B_k(\theta_k)$,
\begin{align}
\Pr\{D_k(\infty) > D^q_{\max}\} = \Pr\{Q_k(\infty) > E^B_k(\theta_k) D^q_{\max}\}.\label{eq:DeqQ}
\end{align}
Then, from \eqref{eq:pdfMD1}, the CCDF of the queueing delay can be derived as
\begin{align}
\Pr\{D_k(\infty) > T_{\rm f}L/s_k(n) \} = \Pr\{Q_k(\infty) > L\} = 1 - \sum\limits_{l = 0}^L {{\pi _l}}\label{eq:CCDFD},
\end{align}
which is too complicated to obtain a
closed-form constraint on queueing delay due to expressions of $\pi_l$ in \eqref{eq:pdfMD1}. Nonetheless, \eqref{eq:CCDFD} can be used to validate the upper bound $P_{D_k}^{\rm UB}$  in \eqref{eq:UB} numerically.


\section{A Framework for Cross-layer Transmission Optimization}
In this section, we first show that the required transmit power to guarantee the queueing delay and transmission error probability requirement for some packets may become unbounded for any given bandwidth and $N_\mathrm{t}$, owing to $D^q_{\max} < T_{\rm c}$. To guarantee the QoS in terms of $D^q_{\max}$ and ${\varepsilon_\mathrm{D}}$ with finite transmit power, we then propose a proactive packet dropping mechanism. Finally, we propose a framework to optimize cross-layer transmission strategy, which includes resource allocation and packet dropping policies  depending on both channel information and queue length.

\subsection{Proactive Packet Dropping and Power Allocation}

We consider the case where $Q_k(n) \geq T_\mathrm{f}E_k^B(\theta_k)$, then $b_k(n) = T_\mathrm{f} E_k^B(\theta_k)$. If a transmit power can guarantee such a departure rate, then for the other case  where $Q_k(n) < T_\mathrm{f}E_k^B(\theta)$, $b_k(n) < T_\mathrm{f}E_k^B(\theta_k)$ can also be supported, i.e., $(D^q_{\max}, {\varepsilon^q_\mathrm{D}})$ can be satisfied according to \eqref{eq:QoSbn}.

Substituting $s_k(n)$ in \eqref{eq:sn} into \eqref{eq:QoS}, we can obtain the required SNR  $\gamma_k$ to ensure $(D^q_{\max}, \varepsilon_k^q)$ and ${\varepsilon_k^c}$ for all packets to the $k$th user using the following equation,
\begin{align}
\ln \left( {1 + \gamma_k} \right) \approx \frac{{T_\mathrm{f} u\ln 2}}{{{\phi}{B N^{\rm c}_k}}}E_k^B(\theta_k) + \sqrt {\frac{V_k}{{{\phi}{B N^{\rm c}_k}}}} f_{\rm Q}^{ - 1}\left( {\varepsilon_k^c} \right). \label{eq:power}
\end{align}

Since ${\bf{h}}_k \sim {\mathbb{C}}^{N_{\rm t}}$ is with i.i.d. elements, the channel gain $g_k = {\bf{h}}_k^H{\bf{h}}_k$ follows Wishart distribution \cite{Telatar1995Capacity}, whose probability density function is
$f_{\rm g}\left( x \right) = \frac{1}{{\left( {{N_\mathrm{t}} - 1} \right)!}}{x^{{N_\mathrm{t}} - 1}}{e^{ - x}}$.
In the considered typical application scenario with $D^q_{\max} < T_{\rm c}$, some packets to be transmitted within the delay bound may experience deep fading with channel gain $g_k$ arbitrarily close to zero.\footnote{This is true also for other channel distribution, say Nakagami-$m$ fading, which is a general model of wireless channels \cite{WirelessCom}.} Then, the required transmit power  to achieve $\gamma_k$ in the $n$th frame, $P_k(n) \triangleq \frac{{{N_0}B N^{\rm c}_k\gamma_k }}{{\mu_k g_k}}$, is unbounded. This means that $s_k(n)$ cannot exceed $E_k^B(\theta_k)$ with finite transmit power if the $n$th frame is in a coherence interval subject to deep fading, even when there is spatial diversity. In other words, for the packets in such an interval, $\varepsilon_k^q+{\varepsilon_k^c}$ will exceed $\varepsilon_D$ will happen if $P_k(n)$ is finite.

To satisfy the QoS requirement with a finite transmit power, we introduce a proactive packet dropping mechanism. By ``proactive", we mean that a packet will be intentionally discarded even when its queueing delay has not exceeded $D^q_{\max}$ in the case $\varepsilon_k^q+{\varepsilon_k^c}>\varepsilon_D$, and then the total number of packets proactively and reactively  dropped\footnote{By ``reactive", we mean that a packet is lost when $D^q_{\max}$ is violated or a coding block is not decoded successfully.} is judiciously controlled to ensure the overall reliability for each user. The rational behind such a mechanism lies in the fact that we only need to ensure the overall packet loss probability ${\varepsilon_D}$ no matter how the packets are lost.

Denote the maximal transmit power of the BS as $P^{\max}$. We discard some packets before transmission in deep fading channels when the required SNR $\gamma_k$ cannot be achieved with $\sum\limits_{k = 1}^K {{P_k}( n )} \leq P^{\max }$. However, we can hardly control the packet dropping probability of each user from $\sum\limits_{k = 1}^K {\frac{{{N_0}{B N^{\rm c}_k}\gamma_k}}{{{\mu_k}g_k}}} \leq P^{\max}$ since the required total transmit power depends on the channel gains of multiple users. To control the packet dropping probability of each user, we introduce the maximal transmit power that can be allocated to the $k$th user $P^{\rm th}_k$. When the required transmit power is higher than $P^{\rm th}_k$, the BS transmits packets to the $k$th user with power $P^{\rm th}_k$ and drop several packets in the $n$th frame. Then, the total transmit power of the BS is bounded by $\sum\limits_{k = 1}^K {P_k^{{\rm{th}}}}$.

To ensure $(D^q_{\max}, \varepsilon_k^q)$  and ${\varepsilon_k^c}$, the \emph{power allocation policy} should depend on both channel gain and queueing length, which is,
\begin{align}
&{P}_k(n) \nonumber\\
& = \left\{ {\begin{array}{*{20}{c}}
{{P_k^{\rm th }},\;\;\;\;\;{\rm{if }}\;Q_k(n) \geq T_\mathrm{f}E_k^B(\theta_k),\;g_k < \frac{{{N_0}B N^{\rm c}_k\gamma_k }}{{\mu_k {P_k^{\rm th }}}}}, \\
{\frac{{{N_0}B N^{\rm c}_k\gamma_k }}{{\mu_k g_k}},\;{\rm{if }}\;Q_k(n) \geq T_\mathrm{f}E_k^B(\theta_k),\; g_k > \frac{{{N_0}B N^{\rm c}_k\gamma_k }}{{\mu_k {P_k^{\rm th }}}} .}
\end{array}} \right.\label{eq:Pn}
\end{align}

In the case $Q_k(n) < T_\mathrm{f}E_k^B(\theta_k)$, ${P}_k(n)$ should satisfy $s_k(n) = Q_k(n)$ when $s_k^{\rm th } > Q_k(n)$ or ${P}_k(n) = P_k^{\rm th }$ when $s_k^{\rm th } \leq Q_k(n)$, where $s_k^{\rm th }$ is the number of packets that can be transmitted in the $n$th frame with $P_k(n) = P_k^{\rm th }$. From the approximation in \eqref{eq:sn}, we obtain $s_k^{\rm th }$ as
\begin{align}
s_k^{\rm th } \approx \frac{{ {\phi}{B N^{\rm c}_k}}}{{u\ln 2}}\left\{ {\ln \left[ {1 + \frac{{{\mu_k}P_k^{\rm th }{g_k}}}{{{N_0}{B N^{\rm c}_k}}}} \right] - \sqrt {\frac{V_k}{{{\phi}{B N^{\rm c}_k}}}} f_{\rm Q}^{ - 1}\left( {{\varepsilon_k^c}} \right)} \right\}\label{eq:sth}.
\end{align}

When ${g}_k < \frac{{{N_0}{B N^{\rm c}_k}\gamma_k}}{{{\mu_k}P_k^{\rm th }}}$ in the $n$th frame, $s_k^{\rm th } < T_\mathrm{f} E_k^B(\theta_k)$. Since  $b_k(n) = \min\{Q_k(n),T_\mathrm{f} E_k^B(\theta_k)\}$ needs to be satisfied to ensure  $(D^q_{\max}, {\varepsilon_k^q})$, the BS has to discard some packets waiting in the queue. Denote the number of packets dropped in the $n$th frame as $b_k^{d}(n) = \max\{b_k(n) - s_k^{\rm th }, 0 \}$.

Then, the \emph{proactive packet dropping policy} is
\begin{align}
b_k^d\left( n \right) = \left\{ {\begin{array}{*{20}{c}}
{\max \left( {{T_\mathrm{f}}E_k^B({\theta}_k) - s_k^{\rm th },0} \right),{\rm{if }}\;{Q}_k\left( n \right) \geq {T_\mathrm{f}}E_k^B({\theta}_k),}\\
{\max \left( {{Q}_k\left( n \right) - s_k^{\rm th },0} \right),\;\;\;{\rm{ if}}\;{{{Q}_k\left( n \right) < {T_\mathrm{f}}E_k^B({\theta}_k).}}}
\end{array}} \right. \label{eq:bd}
\end{align}

This policy is implemented as follows. If $Q_k(n) \geq T_\mathrm{f}E_k^B(\theta_k)$ and $g_k < \frac{{{N_0}B N^{\rm c}_k\gamma_k }}{{\mu_k {P_k^{\rm th }}}}$, then $P_k^{\rm th }$ is used to transmit packets and   $b_k^d\left( n \right)$ packets that cannot be conveyed within the $n$th frame with $P_k^{\rm th }$ are dropped, where $P_k^{\rm th }$ and $b_k^d\left( n \right)$ will be optimized in the next subsection. Since the BS simply discards some packets from the buffer if the channel gain is low, such a policy only introduces negligible processing delay due to several operations of comparison.

Similar to the delivery ratio in \cite{Kumar2009QoS}, we define the packet dropping probability as
\begin{align}
\varepsilon_k^h \triangleq \mathop {\lim }\limits_{N \to \infty } \frac{{\sum\limits_{n = 1}^N {b_k^d\left( n \right)} }}{{\sum\limits_{n = 1}^N {\sum\limits_{i \in {\mathcal{A}}_k} {{a_i}\left( n \right)}} }}= \frac{{\mathbb{E}}[b_k^d\left( n \right) ]}{{{\mathbb{E}}\{\sum\limits_{i \in {\mathcal{A}}_k} {{a_i}\left( n \right)}\}}} \label{eq:defineh},
\end{align}
where the second equality is obtained under the assumption that the queueing system is ergodic, and the average on nominator is taken over both channel gain and queue length.

Based on the analysis in Appendix \ref{App:UB}, the packet dropping probability can be approximated by
\begin{align}
{\varepsilon_k^h} \approx \int_0^{\frac{{{N_0}B N^{\rm c}_k \gamma_k}}{{\mu_k {P_k^{\rm th}}}}} {\left[ {1 - \frac{\ln\left(1+\frac{\mu_k P^{\rm th}_k g_k}{N_0 B N^{\rm c}_k}\right)}{\ln\left(1+\gamma_k\right)}} \right]{f_{\rm g}}\left( g \right)dg}  \label{eq:consteh}.
\end{align}

\subsection{A Framework for Cross-layer Transmission Optimization}
With the proactive packet dropping mechanism, the total transmit power is bounded by $\sum\limits_{k = 1}^K {P_k^{{\rm{th}}}}$. To find the minimal resources required to ensure the QoS, we optimize the cross-layer transmission strategy, which includes a transmit power allocation policy $P_k(n)$ and a proactive packet dropping policy $b_k^d(n)$ for single user scenario and also includes a bandwidth allocation policy for multi-user scenario, to minimize $\sum\limits_{k = 1}^K {P_k^{{\rm{th}}}}$ with given total bandwidth of the system.

According to \eqref{eq:Pn}, $P_k(n)$ depends on the values of $\gamma_k$ and $P_k^{\rm th}$. Given the values of $\gamma_k$ and $\varepsilon_k^h$, the minimal value of $P_k^{\rm th}$ can be obtained from \eqref{eq:consteh} by letting the equality hold. Moreover, the required SNR $\gamma_k$ is determined by  $\varepsilon_k ^c$ and $\varepsilon_k ^q$ according to \eqref{eq:power}. Therefore, the power allocation policy and the minimal $P_k^{\rm th}$ are uniquely determined by the values of $\varepsilon_k ^c$, $\varepsilon_k ^q$ and $\varepsilon_k^h$.

According to \eqref{eq:bd}, the number of packets to be dropped  $b_k^d\left( n \right)$ depends on $s_k^{\rm th}$, which can be obtained from \eqref{eq:sth} after $P_k^{\rm th}$ and $\varepsilon_k^c$ are obtained.

This indicates that to optimize the power allocation policy and packet dropping policy that minimize $\sum\limits_{k = 1}^K {P_k^{{\rm{th}}}}$, we only need to control $\varepsilon_k ^q$, $\varepsilon_k ^c$, and $\varepsilon_k ^h$.

For easy exposition, we first consider single user case, and then extend to multi-user scenario.

\subsubsection{Single-user Scenario}
When $K=1$, the index $k$ can be omitted for notational simplicity. We consider the case that $Q(n)>0$. For $Q(n)=0$, no power is allocated, i.e., $P(n)=0$.

The values of $\varepsilon^c$, $\varepsilon^q$, and $\varepsilon^h$ that minimize $P^{\rm th}$ can be obtained from the following problem,
\begin{align}
\mathop {\min }\limits_{\varepsilon ^q,\varepsilon ^c, \varepsilon ^h}& \quad P^{\rm th} \label{eq:epsilon3}\\
\text{s.t.} & ~ \varepsilon^h = \int_0^{\frac{{{N_0}B N^{\rm c} \gamma}}{{\alpha {P^{\rm th}}}}} {\left[ {1 - \frac{\ln\left(1+\frac{\mu P^{\rm th} g}{N_0 B N^{\rm c}}\right)}{\ln\left(1+\gamma\right)}} \right]{f_{\rm g}}\left( g \right)dg}, \label{eq:epsonh}\tag{\theequation a}\\
&\ln \left( {1 + \gamma} \right) = \frac{{T_\mathrm{f} u\ln 2}}{{{\phi}{B N^{\rm c}}}}E^B(\theta) + \sqrt {\frac{V}{{{\phi}{B N^{\rm c}}}}} f_{\rm Q}^{ - 1}\left( {\varepsilon^c} \right),\label{eq:reqSNR} \tag{\theequation b}\\
&  \varepsilon^c + \varepsilon^q+\varepsilon^h \leq \varepsilon_\mathrm{D} \;\text{and}\; \varepsilon^c, \varepsilon^q,\varepsilon^h \in {\mathbb{R}}^+, \label{eq:eD} \tag{\theequation c}
\end{align}
where constraint \eqref{eq:epsonh} and constraint \eqref{eq:reqSNR} are the single-user case of \eqref{eq:power} and \eqref{eq:consteh}, respectively, $E^B(\theta)$ depends on the source as well as $(D^q_{\max}, \varepsilon^q)$, and ${\mathbb{R}}^+$ represents the positive real number.\footnote{The distribution of channel gain ${f_{\rm g}}\left( g \right)$ depends on the number of antennas  $N_{\rm t}$. Therefore, the optimal solution of problem \eqref{eq:epsilon3} will depend on  $N_{\rm t}$. We will illustrate the impact of $N_{\rm t}$ via numerical results in the next section.}

In the following, we propose a two-step method to find the optimal solution of problem \eqref{eq:epsilon3}.

In the first step, $\varepsilon_{\rm 0}^h \in (0,\varepsilon_\mathrm{D})$ is fixed. Given $\varepsilon_{\rm 0}^h$, $P^{\rm th}$ in the right hand side of \eqref{eq:epsonh} increases with $\gamma$. Hence, minimizing $P^{\rm th}$ is equivalent to minimizing $\gamma$.

For Poisson process, the optimal values of $\varepsilon^c$ and $\varepsilon^q$ that minimize the required $\gamma$ can be obtained by solving the following problem,
\begin{align}
\mathop {\min }\limits_{\varepsilon ^q,\varepsilon ^c}& \quad \frac{{T_\mathrm{f} u\ln 2\ln \left( {1/\varepsilon ^q} \right)}}{{{\phi}{B N^{\rm c}}D_{\max }^q\ln \left[ {1 + \frac{T_\mathrm{f}{\ln \left( {1/\varepsilon^q} \right)}}{{D_{\max }^q{\lambda}}}} \right]}} + \sqrt {\frac{V}{{{\phi}{B N^{\rm c}}}}} f_{\rm Q}^{ - 1}\left( {\varepsilon^c} \right) \label{eq:combine2}\\
\text{s.t.} & \quad  \varepsilon^c + \varepsilon^q \leq \varepsilon_\mathrm{D}-\varepsilon_0^h, \label{eq:relia2}\tag{\theequation a}
\end{align}
where the effective bandwidth in \eqref{eq:EBDepson} is used to derive the objective function.
As proved in Appendix \ref{App:combine}, the objective function in \eqref{eq:combine2} is strictly convex in $\varepsilon^c$ and $\varepsilon^q$, and hence the problem is convex. To ensure the stringent QoS requirement, the required SNR $\gamma$ is high, in this case  $V \approx 1$ as shown in \eqref{eq:dispersion}. Then, there is a unique solution of $\varepsilon^c$ and $\varepsilon^q$ that minimizes $\gamma$. Denote the minimal SNR obtained from problem \eqref{eq:combine2} as $\gamma^*$. Since the right hand side of \eqref{eq:epsonh} decreases with $P^{\rm th}$, for given $\varepsilon _0^h$ and $\gamma^*$, the value of $P^{\rm th}$ can be obtained numerically via binary searching \cite{boyd} as a function of $\varepsilon _0^h$, denoted as $P^{\rm th}(\varepsilon _0^h)$.

In the second step, we find the optimal $\varepsilon_0^h \in (0,\varepsilon_\mathrm{D})$ that minimizes $P^{\rm th}(\varepsilon _0^h)$. Since there is no closed-form expression of $P^{\rm th}(\varepsilon _0^h)$, exhaustive searching is needed to obtain the optimal $\varepsilon_0^h$ in general. However, numerical results indicate that $P^{\rm th}(\varepsilon_0^h)$ first decreases and then increases with $\varepsilon_0^h$. With this property, we can find the optimal solution of $\varepsilon_0^h$ and the required transmit power to ensure $\varepsilon_\mathrm{D}$ via the exact
linear search method \cite{boyd}.

As proved in Appendix \ref{App:optimal}, the solution obtained from the two-step method is the global optimal solution of problem \eqref{eq:epsilon3} if the solutions of both steps are global optimal.


\emph{Impact of traffic feature:} To show the impact of burstiness on the cross-layer optimization, we consider IPP with fixed average packet rate in two asymptotic cases, i.e. $C^2 \to 1$ and $C^2 \to \infty$, where
$C^2$ is the variance coefficient that can be used to characterize burstiness \cite{Jian2015IPP}. To show the impact of burstiness, we keep the average packet rate of IPP, $\frac{\alpha}{\alpha+\beta}\lambda^{\rm on}$, as a constant. Then, the average packet rate can be expressed as $\frac{\lambda^{\rm on}}{1+\delta}$, and $C^2 = 1 +\frac{2\delta\lambda^{\rm on}}{(1+\delta)^2\alpha}$ \cite{Jian2015IPP}, where $\delta = \beta/\alpha $.

When $\alpha \to \infty$, $C^2 \to 1$, the effective bandwidth of the IPP can be expressed as
$E^{B}(\theta) = \frac{{{\lambda^{\rm on}}}}{{T_{\rm f}{\theta}}(1+\delta)}\left( {{e^{{\theta}}} - 1} \right)$,
which is the same as the effective bandwidth of a Poisson process with average packet rate $\frac{\lambda^{\rm on}}{1+\delta}$. When $\alpha \to 0$, $C^2 \to \infty$, the effective bandwidth of the IPP can be expressed as
$E^{B}(\theta) = \frac{{{\lambda^{\rm on}}}}{{T_{\rm f}{\theta}}}\left( {{e^{{\theta}}} - 1} \right)$, which is the same as the effective bandwidth of a Poisson process with average packet rate $\lambda^{\rm on}$.

To show the impact of autocorrelation, we consider a SPP with parameters $(\lambda^{\rm I},\lambda^{\rm II},\alpha_{\rm I},\alpha_{\rm II})$, where $\lambda^{\rm I} \in [0,\lambda^{\rm on}]$, $\lambda^{\rm II} = \lambda^{\rm on}$, $\alpha_{\rm I} = \alpha$ and $\alpha_{\rm II} = \beta$. An upper bound of the effective bandwidth of it can be obtained by substituting $\lambda = \lambda^{\rm on}$ into \eqref{eq:EBPoisson}. Therefore, the effective bandwidth of SPP is less than that of a Poisson process with average packet rate $\max\{\lambda^{\rm I},\lambda^{\rm II}\}$.

{\bf{Remark 2:}} For IPP, when $C^2$ increases from $1$ to $\infty$, the effective bandwidth (i.e. the required constant service rate) increases $1+\delta$ times. For SPP, the required constant service rate does not exceed the upper bound, which equals to the effective bandwidth of a Poisson process with average packet rate $\max\{\lambda^{\rm I},\lambda^{\rm II}\}$. This indicates that the service rate requirement is still finite for IPP with $C^2\to \infty$ or for SPP with any values of $\alpha_{\rm I}$ and $\alpha_{\rm II}$. Therefore, the burstiness and autocorrelation will not change the proposed framework.

\subsubsection{Multi-user Scenario}
In this case, we jointly optimize $N^{\rm c}_k$, $\varepsilon_k^c$, $\varepsilon_k^q$, and $\varepsilon_k^h$, with which we can obtain the optimal cross-layer strategy including bandwidth allocation, power allocation and packet dropping policies. The optimization problem in the multi-user scenario is formulated as
\begin{align}
&\mathop {\mathop {\min }\limits_{{N^{\rm c}_k},\varepsilon _k^q,\varepsilon _k^c,\varepsilon _k^h} }\limits_{k = 1,2,...,K} P^{\rm tot} \triangleq \sum\limits_{k = 1}^K {P_k^{\rm th }}  \label{eq:minPt}\\
&\text{s.t.}\; \varepsilon_k^h = \int_0^{\frac{{{N_0}B N^{\rm c}_k \gamma_k}}{{\mu_k {P_k^{\rm th}}}}} {\left[ {1 - \frac{\ln\left(1+\frac{\mu_k P^{\rm th}_k g}{N_0 B N^{\rm c}_k}\right)}{\ln\left(1+\gamma_k\right)}} \right]{f_{\rm g}}\left( g \right)dg},\label{eq:epsonhk}\tag{\theequation a}\\
&\ln \left( {1 + \gamma_k} \right) = \frac{{T_\mathrm{f} u\ln 2}}{{{\phi}{B N^{\rm c}_k}}}E_k^B(\theta_k) + \sqrt {\frac{V_k}{{{\phi}{B N^{\rm c}_k}}}} f_{\rm Q}^{ - 1}\left( {\varepsilon_k^c} \right),\label{eq:reqSNRk} \tag{\theequation b}\\
&\varepsilon_k^c+\varepsilon_k^q+\varepsilon_k^h \leq \varepsilon_\mathrm{D} \;\text{and}\; \varepsilon_k^c,\varepsilon_k^q,\varepsilon_k^h \in {\mathbb{R}}^+,\label{eq:eDK}\tag{\theequation c}\\
&\sum\limits_k^K N^{\rm c}_k  \le N^{\rm c}_{\max}, N^{\rm c}_k \in {\mathbb{Z}}^+ , k=1,...,K, \label{eq:Wk}\tag{\theequation d}
\end{align}
where $N^{\rm c}_{\max}$ is the maximal number of subcarriers for DL transmission.\footnote{By solving problem \eqref{eq:minPt}, the bandwidth  (i.e., the number of subcarriers) allocation is obtained. With constraint \eqref{eq:Wk}, the total number of subcarriers allocated to all the $K$ users is less than the
maximal number of subcarriers of the system. Therefore, we can always find a subcarrier allocation policy, with which each subcarrier is only
allocated to one user.} Since $N^{\rm c}_k$ is integer, this is a mixed-integer programming problem.

Given the values of $N^{\rm c}_k, k = 1,...,K$, the problem can be decomposed into $K$ single-user problems similar to \eqref{eq:epsilon3}, which can be solved by the two-step method. Then, the power allocation policy among subsequent TTIs and the packet dropping policy can be obtained similarly to those in the single-user scenario, i.e., \eqref{eq:Pn} and \eqref{eq:bd}.
We refer to the $K$ single-user problems as \emph{subproblem I}. Since binary search and exact linear search methods are applied in solving subproblem I, the complexity of the two-step method is $O(\log_2(\frac{\varepsilon_{\rm D}}{\Delta^h})\log_2(\frac{\varepsilon_{\rm D}}{\Delta^c}))$.\footnote{The complexity of a searching algorithm depends on the stopping criterion. Here, the iterations stop if $|\varepsilon_k^h(i)-\varepsilon_k^{h}(i+1)|<\Delta^h$ or $|\varepsilon_k^c(i)-\varepsilon_k^{c}(i+1)|<\Delta^c$ is satisfied, where $\varepsilon_k^h(i)$ and $\varepsilon_k^c(i)$ are the results obtained after $i$ iterations.}

The complexity of problem \eqref{eq:minPt} is determined by the integer programming that optimizes $N^{\rm c}_k, k = 1,...,K$ with given $\varepsilon_k^c,\varepsilon_k^q,\varepsilon_k^h$ to minimize the objective function in \eqref{eq:minPt}. We refer this integer programming as \emph{subproblem II}. Since $N^{\rm c}_k \geq 1$, the remaining number of subcarriers is $N^{\rm c}_{\max} - K$. To solve problem \eqref{eq:minPt}, we need to allocate the remaining subcarriers to $K$ users. Thus, subproblem II includes around $K^{N^{\rm c}_{\max}-K}$ feasible solutions. To reduce complexity, a heuristic algorithm is proposed, as listed in Table II. The basic idea is similar to the steepest descent method \cite{boyd}. The subcarrier allocation algorithm includes $N^{\rm c}_{\max}-K$ steps. In each step, one subcarrier is allocated to one of the $K$ users that leads to the steepest total transmit power descent. The proposed algorithm only needs to solve subproblem I for $K{( N^{\rm c}_{\max}-K)}$ times, and hence the complexity is $O\left(K{( N^{\rm c}_{\max}-K)}\right)$. Further considering the complexity of the two-step method for solving subproblem I, the overall complexity of the proposed algorithm is $O\left(K{( N^{\rm c}_{\max}-K)}\log_2(\frac{\varepsilon_{\rm D}}{\Delta^h})\log_2(\frac{\varepsilon_{\rm D}}{\Delta^c})\right)$.

\renewcommand{\algorithmicrequire}{\textbf{Input:}}
\renewcommand{\algorithmicensure}{\textbf{Output:}}
\begin{table}[htb]\small
\caption{Subcarrier Allocation Algorithm}
\vspace{-0.3cm}
\begin{tabular}{p{8.5cm}}
\\\hline
\end{tabular}
\vspace{-0.3cm}
\begin{algorithmic}[1]
\REQUIRE Number of users $K$, total number of subcarrers $N^{\rm c}_{\max}$, duration for data transmission in each DL frame $\phi$, packet size $u$, noise spectral density $N_0$, number of transmit antennas $N_\mathrm{t}$, average channel gains of users $\mu_k$, $k=1,...,K$.
\ENSURE Subcarrier allocation $N^{\rm c^*}_k$, $k=1,...,K$.
\STATE Set $N^{\rm c}_k(0) := 1$, $k=1,...,K$. Set $l := 1$.
\STATE Solve subproblem I with $N^{\rm c}_k(0) = 1$, and obtain the total transmit power $P^{\rm tot}{(0)}$.
\WHILE{$l \leq N^{\rm c}_{\max} - K $}
\STATE Set $\hat{k} := 1$
\WHILE{$\hat{k} \leq K$}
\STATE $N^{\rm c}_{\hat{k}}{(l)} := N^{\rm c}_{\hat{k}}(l-1) + 1$; $N^{\rm c}_{{k}}{(l)} := N^{\rm c}_{{k}}{(l-1)}$, $k \ne \hat{k}$.
\STATE Solve subproblem I with $N^{\rm c}_k{(l)}$, and obtain $\hat{P}_{\hat{k}}^{\rm tot}{(l)}$.
\STATE $\hat{k} := \hat{k}+1$.
\ENDWHILE
\STATE $k^* := \arg \mathop {\min }\limits_{\hat k} \hat P_{\hat k}^{\rm tot}{(l)}$.
\STATE $N^{\rm c}_{{k}^*}{(l)} := N^{\rm c}_{{k}^*}{(l-1)} + 1$; $N^{\rm c}_{{k}}{(l)} := N^{\rm c}_{{k}}{(l-1)}$, $k \ne {k}^*$.
\STATE $l := l + 1$.
\ENDWHILE
\RETURN $N^{\rm c^*}_k = N^{\rm c}_{{k}}{(l-1)}, k=1,...,K$.
\end{algorithmic}
\vspace{-0.2cm}
\begin{tabular}{p{8.5cm}}
\\
\hline
\end{tabular}
\vspace{-0.2cm}
\end{table}

\section{Applying the Framework to Frequency-selective Channel}
If  the number of users is not very large, the bandwidth allocated to a user (say $W_k = B N^{\rm c}_k$ in problem \eqref{eq:minPt}) could be larger than the coherence bandwidth. In this section, we show how to apply the framework to frequency-selective channel.

We divide the bandwidth allocated  to the $k$th user into  $N^{\rm sc}_k$ subchannels, where each subchannel consists of multiple subcarriers. The bandwidth of each subchannel is $W_{\rm c}$ that is less than the coherence bandwidth. Then, the subcarriers within each subchannel subject to flat fading, while the subchannels subject to frequency-selective fading. To study the delay and reliability performance, we first need to find the achievable rate with finite blocklength. As shown in Appendix \ref{App:Nschannel}, the number of packet that can be transmitted in one frame can be obtained as,
\begin{align}
s^{\rm fs}_k \approx \frac{{{\phi}{W_{\rm c}}}}{{u\ln 2}}&\left\{\sum\limits_{j = 1}^{{N^{\rm sc}_k}} {\ln \left[ {1 + \frac{{{\mu_k}P_{kj}(n) {g_{kj}}}}{{{N_0}{W_{\rm c}}}}} \right]}  - \sqrt {\frac{V_k}{{{\phi}{W_{\rm c}}}}} f_{\rm Q}^{ - 1}\left( {\varepsilon_k^c} \right)\right\}\nonumber\\
&\quad\quad\quad\quad\quad\quad\quad\quad\quad\quad\quad\quad\text{(packets)},\label{eq:Nkst}
\end{align}
where $P_{kj}(n)$ is the transmit power allocated to the $j$th subchannel of the $k$th user in the $n$th frame, $g_{kj}$ is the instantaneous channel gain on the $j$th subchannel of the $k$th user, and $V_k = {N^{\rm sc}_k} - \sum\limits_{j = 1}^{{N^{\rm sc}_k}} {\frac{1}{{{{\left[ {1 + \frac{{\mu_k} P_{kj}(n){g_{kj}}}{{N_0}{W_{\rm c}}}} \right]}^2}}}}$. Since the channel gains could be arbitrarily close to zero, the required transmit power to guarantee queueing delay is also unbounded.

\begin{figure}[htbp]
        \vspace{-0.2cm}
        \centering
        \begin{minipage}[t]{0.45\textwidth}
        \includegraphics[width=1\textwidth]{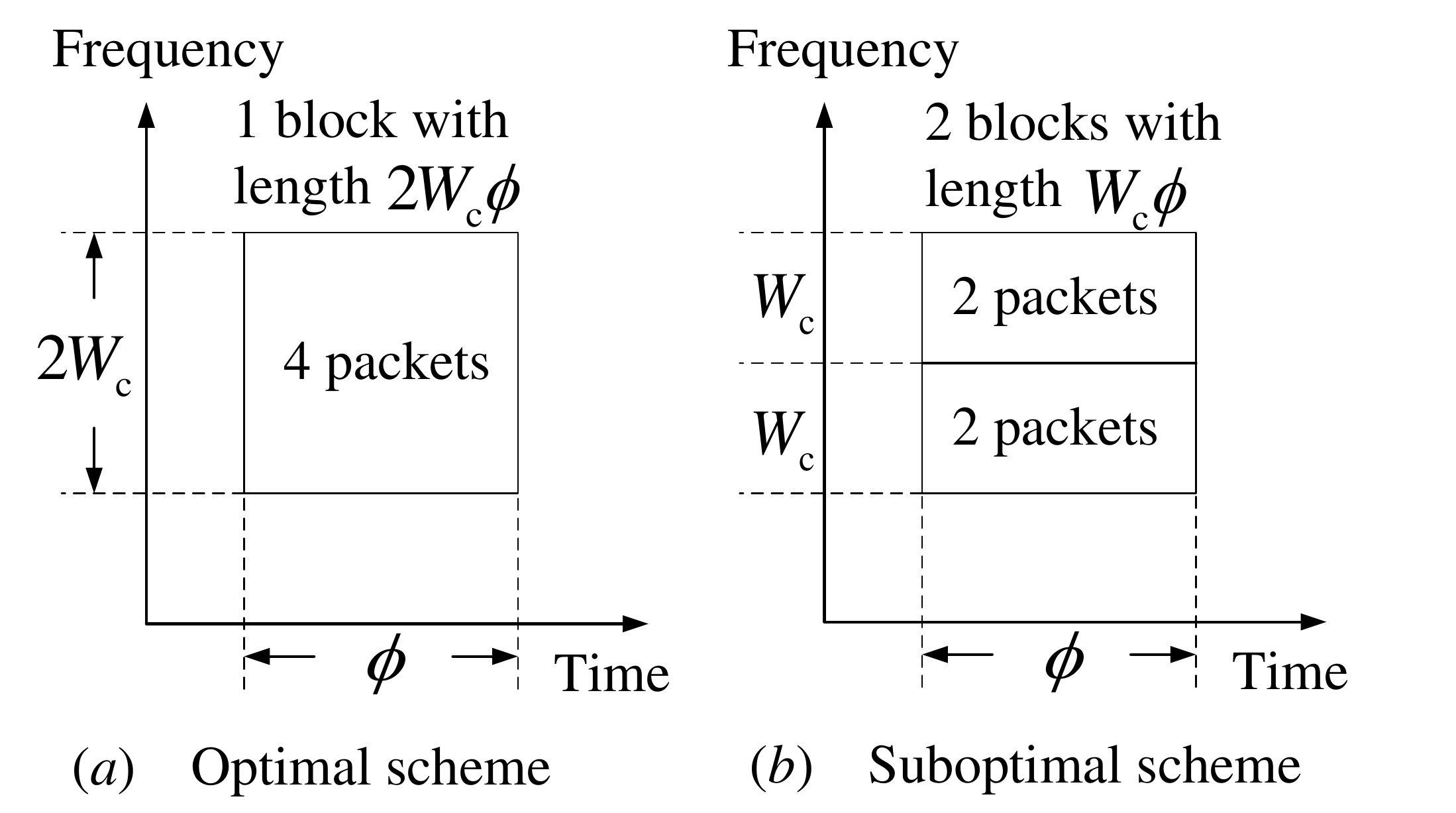}
        \end{minipage}
        \vspace{-0.3cm}
        \caption{Illustration of two channel coding schemes, where four packets need to be transmitted in a frame and $W_k=2 W_{\rm c}$.}
        \label{fig:coding}
        \vspace{-0.2cm}
\end{figure}

The packet rate in \eqref{eq:Nkst} can be achieved if all the packets in a frame are coded in one block with length $W_{\rm c}N^{\rm sc}_k\phi$ (called the optimal coding scheme), as illustrated in Fig. \ref{fig:coding}(a). By substituting \eqref{eq:Nkst} into \eqref{eq:QoS}, we cannot obtain the required SNR to ensure $(D^q_{\max}, \varepsilon_k^q)$ and ${\varepsilon_k^c}$ as that in \eqref{eq:power}. This is because each channel coding block consists of packets transmitted over multiple subchannels with different instantaneous channel gains. As a result, it is very challenging to  derive and optimize the proactive packet dropping probability that guarantees the QoS.

To overcome this difficulty, we consider a suboptimal coding scheme that the packets to be transmitted on different subchannels are coded independently.  As illustrated in Fig. \ref{fig:coding}(b), the blocklength of the suboptimal coding scheme is $W_{\rm c}\phi$. With shorter blocklength, the suboptimal coding scheme can support lower packet rate for a given ${\varepsilon_k^c}$, thus the required resources with the suboptimal channel coding scheme are higher than that with the optimal scheme in order to achieve the same QoS \cite{Giuseppe2016Toward}. Nonetheless, with the optimal scheme, if a block is not decoded without error, then all the packets transmitted in one frame will be lost. By contrast, with the suboptimal scheme, if the packets in one block is not decoded successfully, the packets in other blocks can still be decoded correctly. This suggests that the packet transmission errors with the suboptimal scheme is less busty than those with the optimal scheme.\footnote{Some applications like safe messages transmission in vehicle networks may prefer such suboptimal scheme, which is also applicable for flat fading channels.}

When the number of packets transmitted over each subchannel is $E_k^B(\theta_k)/N^{\rm sc}_k$, the constraints on proactive packet dropping probability, queueing delay violation probability and transmission error probability can be obtained by replacing $B N^{\rm c}_k$ and $E_k^B(\theta_k)$ in \eqref{eq:epsonhk} and \eqref{eq:reqSNRk} with $W_{\rm c}$ and $E_k^B(\theta_k)/N^{\rm sc}_k$, respectively. In this way, the proposed framework can be applied over frequency-selective channel.

In what follows, we analyze the rate loss. With the suboptimal scheme, the number of packets that can be transmitted over the ${N^{\rm sc}_k}$ subchannels can be expressed as follows,
\begin{align}
\tilde s_k^{{\rm{fs}}} \approx \frac{{\phi {W_{\rm{c}}}}}{{u\ln 2}}&\sum\limits_{j = 1}^{N_k^{{\rm{sc}}}} {\left\{ {\ln \left[ {1 + \frac{{{\mu _k}{P_{kj}}(n){g_{kj}}}}{{{N_0}{W_{\rm{c}}}}}} \right] - \sqrt {\frac{{{{\tilde V}_{kj}^{{\rm{fs}}}}}}{{\phi {W_{\rm{c}}}}}} f_{\rm{Q}}^{ - 1}\left( {\varepsilon _k^c} \right)} \right\}} \nonumber\\
&\quad\quad\quad\quad\quad\quad\quad\quad\quad\quad\quad\quad\text{(packets)},\label{eq:codingst}
\end{align}
where the number of packets transmitted in each subchannel is obtained by replacing bandwith $BN_k^{\rm c}$ in \eqref{eq:sn} with $W_{\rm c}$, and hence ${\tilde V}_{kj}^{\rm{fs}} = 1-\frac{1}{\left[1+\frac{\mu_k P _{kj}(n)g_{kj}}{ N_0 W_{\rm c}}\right]^2}$. From \eqref{eq:Nkst} and \eqref{eq:codingst}, we can derive the gap between $s^{\rm fs}_k$ and $\tilde s_k^{{\rm{fs}}}$ as,
\begin{align}
s^{\rm fs}_k - \tilde s_k^{{\rm{fs}}} \approx \frac{{{{\sqrt {\phi W} }_{\rm{c}}}}}{{u\ln 2}}\left( {\sum\limits_{j = 1}^{N_k^{{\rm{sc}}}} {\sqrt {{\tilde V}_{kj}^{\rm{fs}}} }  - \sqrt {V_k} } \right)f_{\rm{Q}}^{ - 1}\left( {\varepsilon _k^c} \right),\nonumber
\end{align}
which shows that $s^{\rm fs}_k - \tilde s_k^{{\rm{fs}}} \sim O(N_k^{{\rm{sc}}} - \sqrt{N_k^{{\rm{sc}}}})$, \footnote{Here $y(N_k^{{\rm{sc}}}) \sim O\left(x(N_k^{{\rm{sc}}})\right)$ means $y(N_k^{{\rm{sc}}})/x(N_k^{{\rm{sc}}})$ approaches to a constant when $N_k^{{\rm{sc}}}$ is large.} and thus the gap between $s^{\rm fs}_k$ and $\tilde s_k^{{\rm{fs}}}$ increases with $N_k^{{\rm{sc}}}$. From \eqref{eq:codingst}, we have $\tilde s_k^{{\rm{fs}}} \sim O(N_k^{{\rm{sc}}})$, hence $({s^{\rm fs}_k - \tilde s_k^{{\rm{fs}}}})/{\tilde s_k^{{\rm{fs}}}} \sim O(1)$. This means that the normalized rate loss $({s^{\rm fs}_k - \tilde s_k^{{\rm{fs}}}})/{\tilde s_k^{{\rm{fs}}}}$ approaches to a constant when $N_k^{{\rm{sc}}}$ is large.

\section{Simulation and Numerical Results}
In this section, we first validate that the effective bandwidth can be used as a tool to optimize resource allocation in short delay regime for Poisson process, IPP and SPP. Then, we show the optimal values of $\varepsilon _k^q$, $\varepsilon _k^c$ and $\varepsilon _k^h$, and the required maximal transmit power  for both Poisson process and IPP.\footnote{The optimal values of $\varepsilon _k^q$, $\varepsilon _k^c$ and $\varepsilon _k^h$ and the required transmit power for SPP are similar to that for IPP, and hence the results for SPP are omitted for conciseness.} Next, we compare the required transmit power of the proposed algorithm with the global optimal policy obtained by exhaustive searching.

A single-BS scenario is considered in the sequel. The users are uniformly distributed with distances from the BS as $50$~m $\sim$ $200$~m. The arrival process of each user is modeled as Poisson process, IPP, or SPP with average rate $1000$~packets/s, i.e., each user requests the safety messages from $50$ nearby sensors, and each sensor uploads packets to the BS with average rate $20$~packets/s \cite{Hassan2013A}. Other parameters are listed in Table III, unless otherwise specified.

\begin{table}[htbp]
\vspace{-0.2cm}\small
\renewcommand{\arraystretch}{1.3}
\caption{Parameters \cite{A2014Scenarios,Hassan2013A}}
\begin{center}\vspace{-0.2cm}
\begin{tabular}{|p{5cm}|p{2.5cm}|}
  \hline
  Overall reliability requirement $\varepsilon _\mathrm{D}$ & $1-99.99999\%$ \\\hline
  E2E delay requirement $D_{\max}$ & $1$~ms \\\hline
  Queueing delay requirement $D^q_{\max}$ & $0.8$~ms \\\hline
  Duration of each frame (equals to TTI) & $0.1$~ms \\\hline
  Duration of data transmission in one frame $\phi$ & $0.06$~ms \\\hline
  Single-sided noise spectral density $N_0$ & $-173$~dBm/Hz \\\hline
  Packet size $u$ & $20$~bytes\\\hline
  Path loss model $10\lg(\mu_k)$ & $35.3+37.6 \lg(d_k)$ \\\hline
  Average duration of ``OFF" or ``ON" state $\alpha^{-1}$ or $\beta^{-1}$ & $1$~s (i.e. $10^{4}$ frames)\\\hline
\end{tabular}
\end{center}
\vspace{-0.5cm}
\end{table}

\begin{figure}[htbp]  
\vspace{-0.2cm}
\centering
\subfigure[{Poisson arrivals, where $\varepsilon_k^q = 10^{-8}$.}]{
\label{fig:subfig:CCDFQ} 
\includegraphics[width=0.45\textwidth]{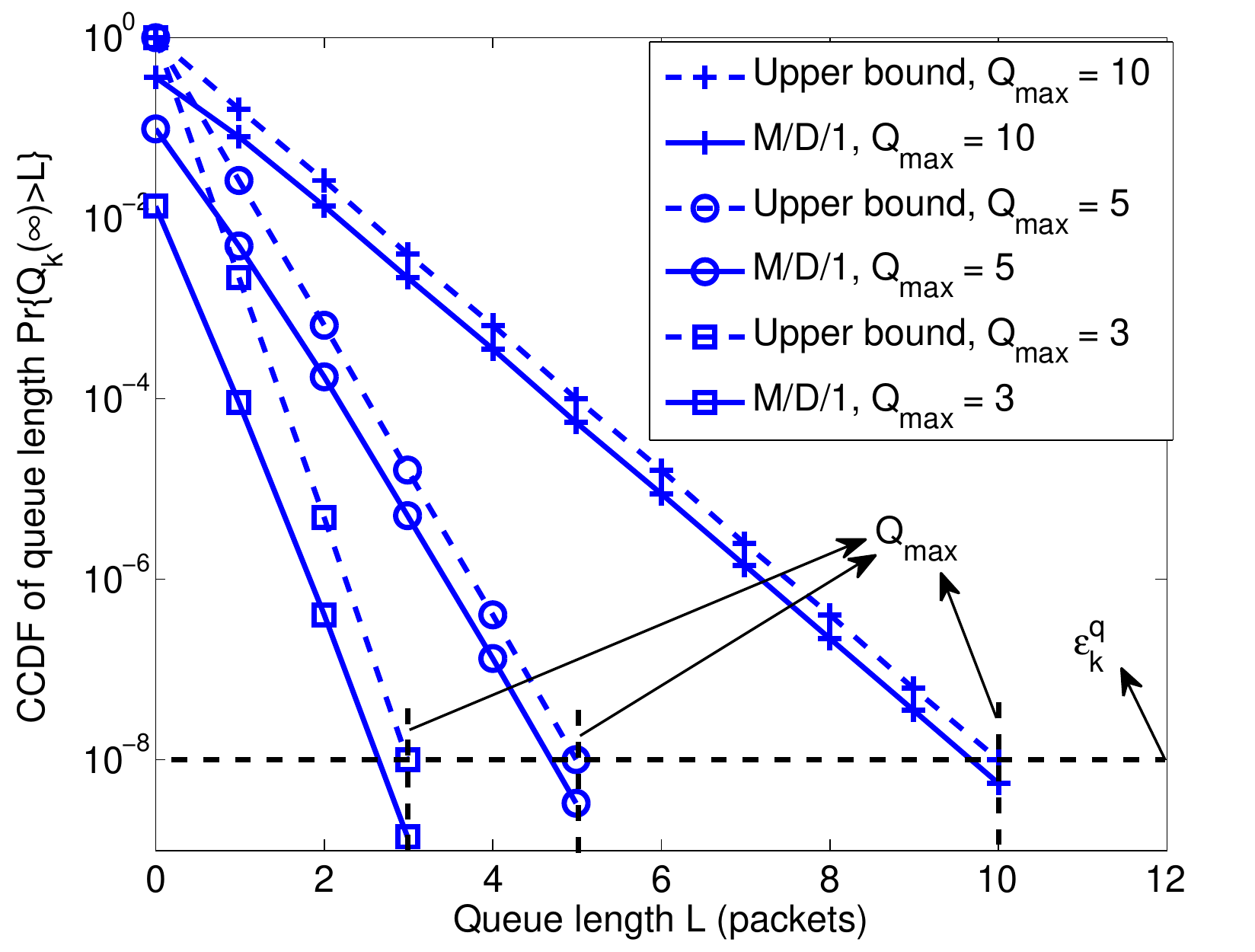}}
\subfigure[{Poisson arrival, IPP and SPP, where $C^2 = 1001$ for the IPP.}]{
\label{fig:subfig:CCDFD} 
\includegraphics[width=0.45\textwidth]{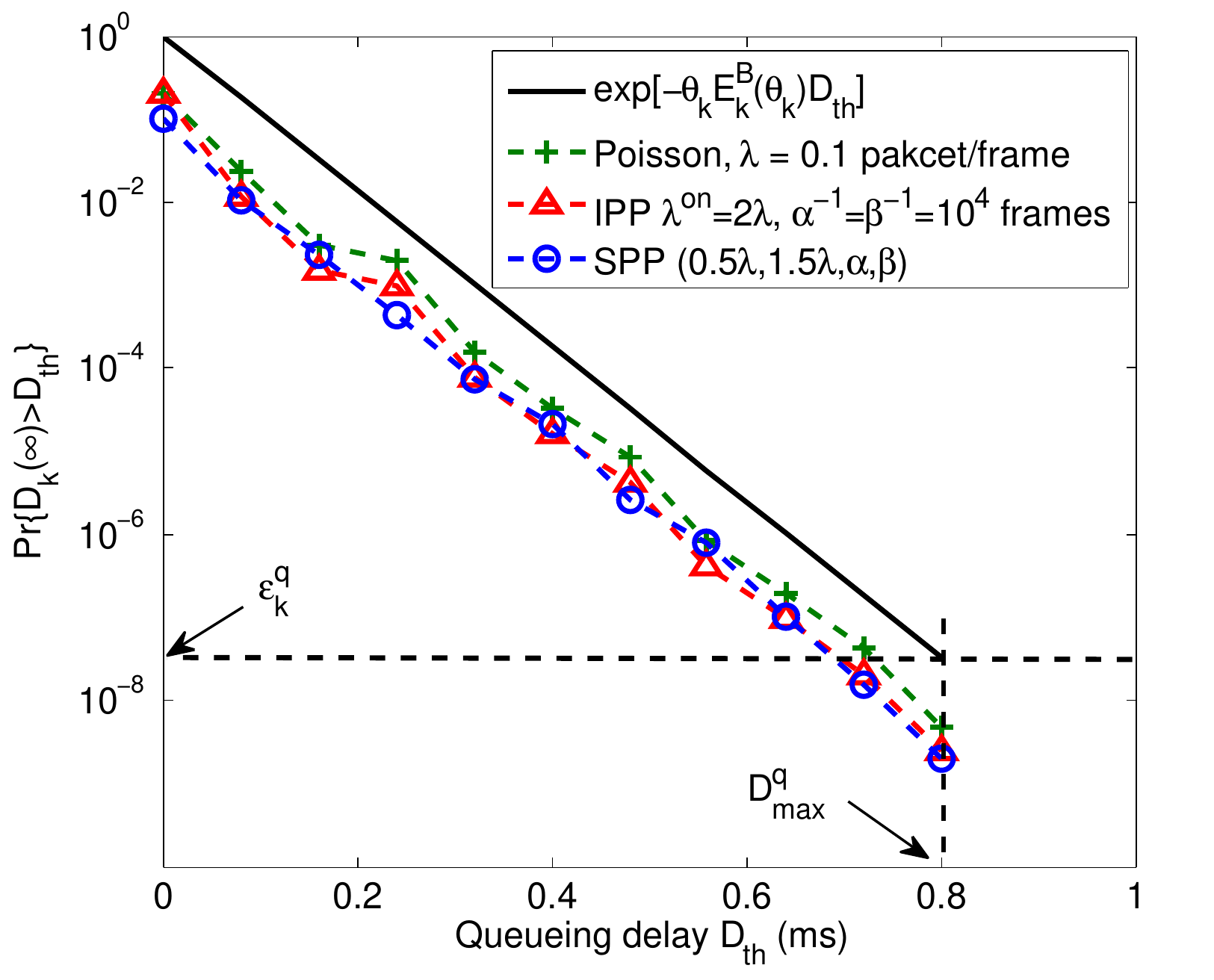}}
\caption{Validating the upper bound in \eqref{eq:UB}.}
 \label{fig:CCDF} 
\vspace{-0.2cm}
\end{figure}
The CCDFs of queue length and queueing delay for the packets to the $k$th user are shown in Fig. \ref{fig:CCDF}, where \eqref{eq:DeqQ} is used to translate the CCDF of the queueing delay into the CCDF of queue length. To obtain the upper bound in \eqref{eq:UB}, $\Pr\{D_k(\infty) > D_{\rm th}\} \leq \exp\{-\theta_k E_k^B(\theta_k) D_{\rm th}\}$ is computed by changing $D_{\rm th} $ from $0$ to $D^q_{\max}$. The CCDFs of queueing delay are obtained via Monte Carlo simulation by generating arrival process and service process during $10^{10}$~frames. Numerical results in Fig. \ref{fig:subfig:CCDFQ} indicate that for Poisson process, the upper bound derived by effective bandwidth works when the maximal queue length is short. Simulation results in Fig. \ref{fig:subfig:CCDFD} show that the upper bound also works for IPP and SPP. In fact, it has been observed in  \cite{ECMIMO} that effective bandwidth can be used for resource allocation under statistical queueing delay requirement when $D_{\max}^q$ is small, if the TTI is much shorter than the delay bound.
\begin{figure}[htbp]  
\vspace{-0.2cm}
\centering
\subfigure[{Optimal values of ${\varepsilon^c}$, ${\varepsilon^q}$ and ${\varepsilon^h}$ that minimize the required transmit power.}]{
\label{fig:subfig:1} 
\includegraphics[width=0.45\textwidth]{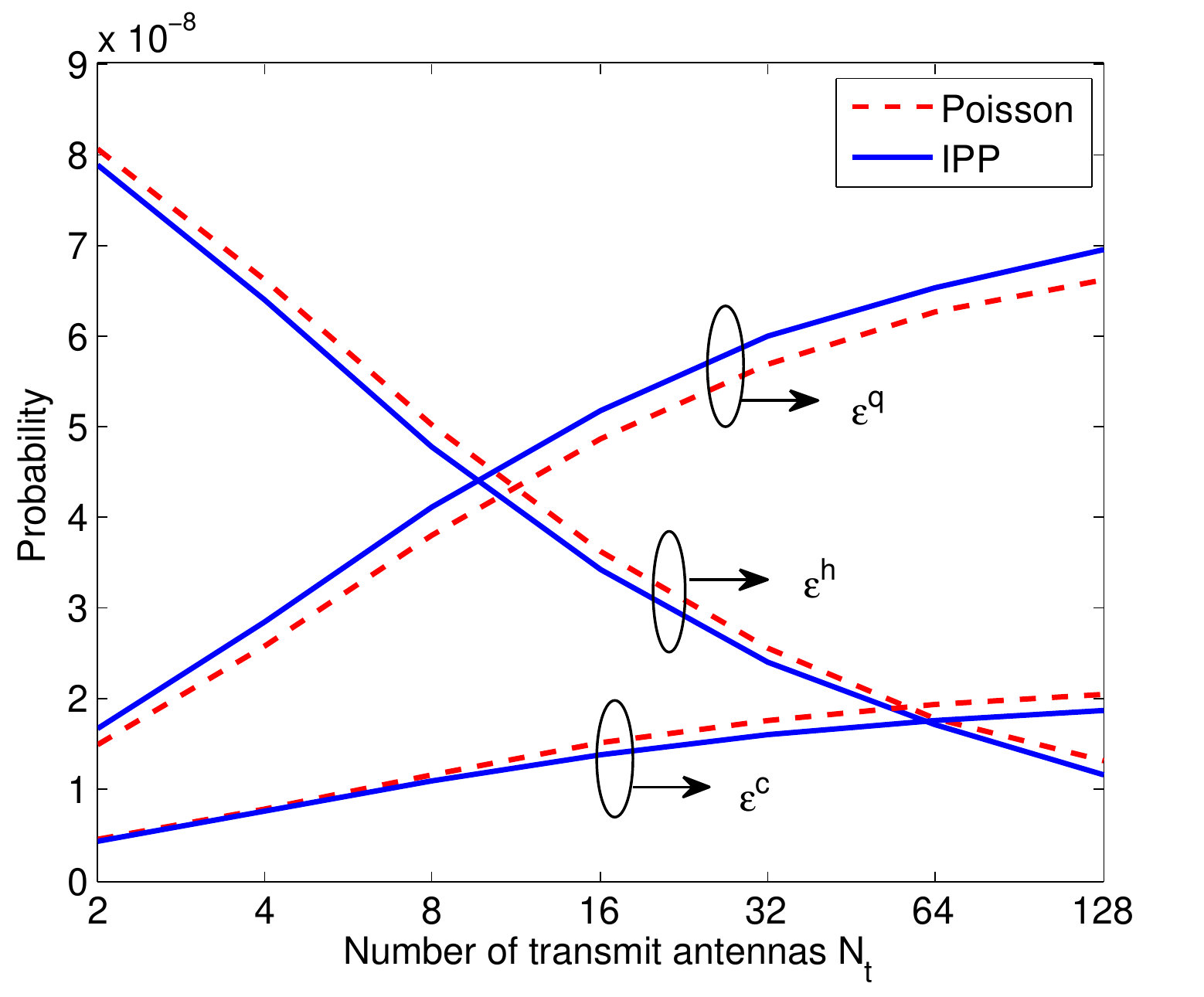}}
\subfigure[{Required maximal transmit power.}]{
\label{fig:subfig:2} 
\includegraphics[width=0.45\textwidth]{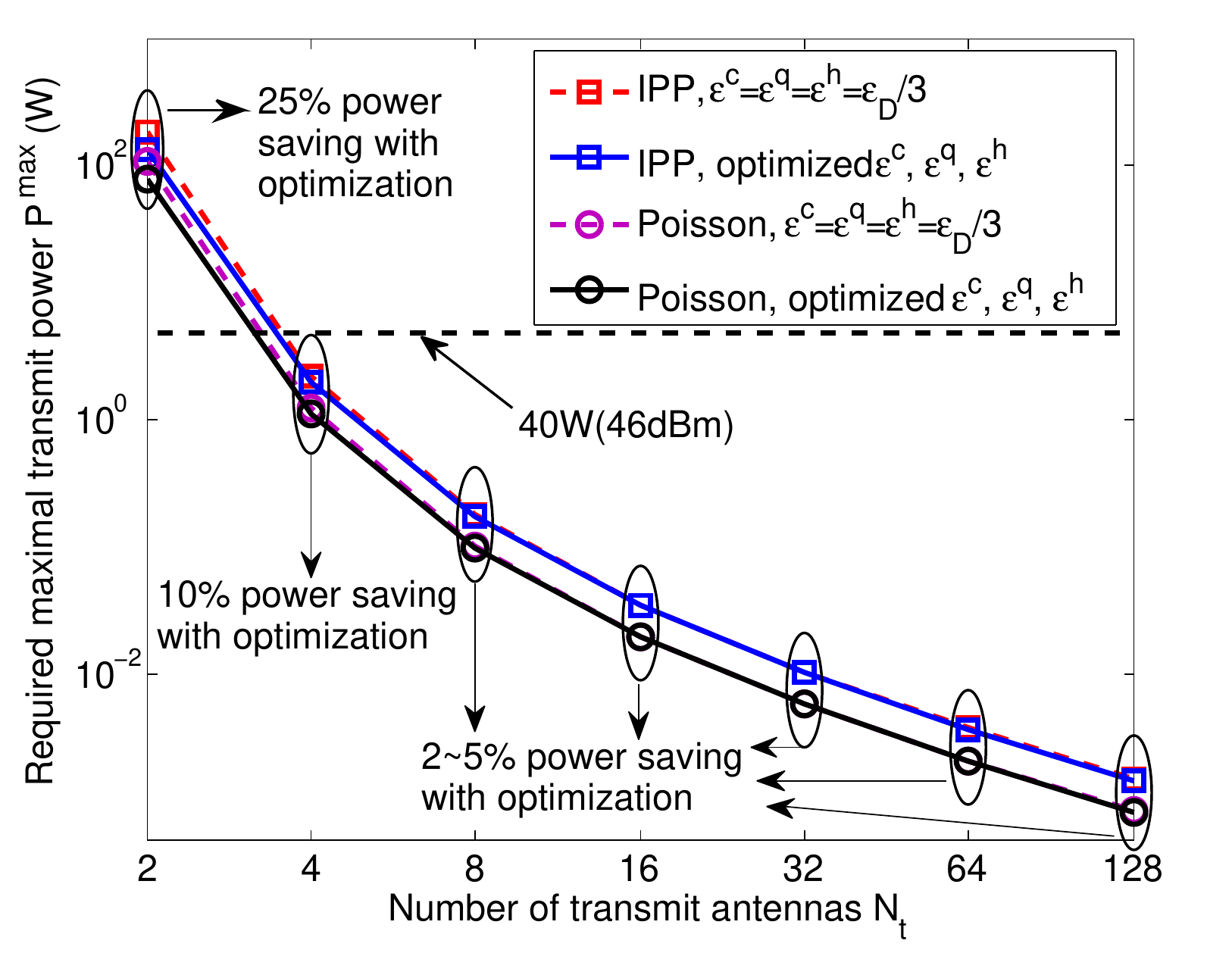}}
\caption{Single-user scenario, where user-BS distance is $200$~m, $N^{\rm c} = 4$, $B = 0.15$~MHz, and $\alpha = \beta$.}
 \label{fig:epsons} 
\vspace{-0.2cm}
\end{figure}

The optimal solution of problem \eqref{eq:epsilon3} and the required maximal transmit power for both Poisson  and IPP are shown in Fig. \ref{fig:epsons}. The results in Fig. \ref{fig:subfig:1} show that ${\varepsilon _k^c}$, ${\varepsilon _k^q}$ and ${\varepsilon _k^h}$ are in the same order of magnitude with different values of $N_{\rm t}$. In fact, similar to ${\varepsilon _k^h}$, when either ${\varepsilon _k^c}$ or ${\varepsilon _k^q}$ is set as zero,  the required transmit power will become infinite, because $E^{B}_k(\theta_k) \to \infty$ when ${\varepsilon _k^q}=0$ (as can be clearly seem from \eqref{eq:EBDepson}) and $f_{\rm Q}^{-1}(x) \to \infty$ (and hence $s_k(n)$ in \eqref{eq:sn} approaches infinity) when ${\varepsilon _k^c}=0$.  This implies that the optimal probabilities will also be in the same order when other system parameters change. On the other hand, Fig. \ref{fig:subfig:2} shows that compared with ${\varepsilon _k^c}={\varepsilon _k^q} = {\varepsilon _k^h}$, the required maximal transmit power only reduces $2\sim5$\%  with the optimized ${\varepsilon _k^c}$, ${\varepsilon _k^q}$ and ${\varepsilon _k^h}$ when $N_{\rm t} \geq 8$. This implies that dividing the required packet loss probability equally to the three probabilities will cause minor performance loss.

Moreover, the optimal queueing delay violation probability for IPP is higher than that for Poisson process. This indicates that bursty arrival processes lead to higher  queueing delay violation probability. Furthermore, $P^{\rm th}$ decreases extremely fast as $N_{\rm t}$ increases. This agrees with the intuition: increasing the number of transmit antennas is an efficient way to reduce the required maximal transmit power thanks to the spatial diversity.

\begin{table}[btp]
\vspace{-0.0cm}
\small
\renewcommand{\arraystretch}{1.3}
\caption{Required Transmit Power, $N^{\rm c}_{\max} = 16$, $B = 0.15$~MHz, and $N_\mathrm{t} = 8$}
\begin{center}\vspace{-0.2cm}
\begin{tabular}{|p{3.0 cm}|p{1.2cm}|p{1.2cm}|p{1.2cm}|}
\hline
  Number of users $K$& 2& 4 & 6 \\\hline
  Proposed Algorithm & $0.155$~W& $0.519$~W& $1.979$~W  \\\hline
  Exhaustive Searching & $0.155$~W& $0.519$~W& $1.979$~W  \\\hline
\end{tabular}
\end{center}
\vspace{-0.8cm}
\end{table}

The required $\sum\limits_{k = 1}^K {P_k^{\rm th }}$ obtained by the proposed algorithm and the global optimal solution with exhaustive searching are provided in Table IV. The results illustrate that the proposed algorithm is near-optimal. Because the complexity of exhaustive search method is extremely high when $N^{\rm c}_{\max}$ and $K$ are large, we only provide results with small values of $N^{\rm c}_{\max}$ and $K$.

\begin{figure}[htbp]
        \vspace{-0.6cm}
        \centering
        \begin{minipage}[t]{0.45\textwidth}
        \includegraphics[width=1\textwidth]{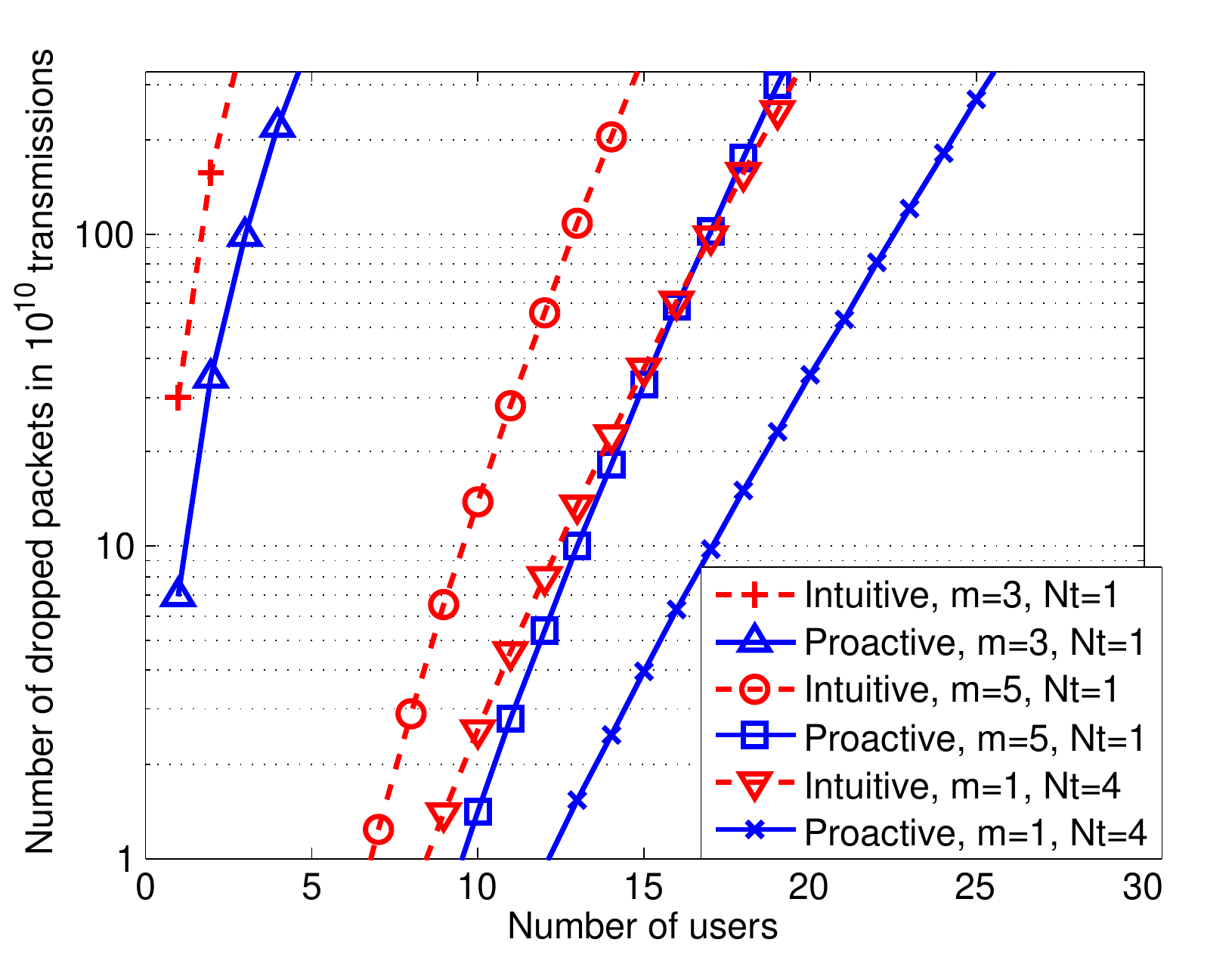}
        \end{minipage}
        \vspace{-0.3cm}
        \caption{Number of dropped packets over Nakagami-$m$ fading channel, where $N^{\rm c}_{\max} = 1024$ and $P^{\max} = 46$~dBm.}
        \label{fig:model}
        \vspace{-0.4cm}
\end{figure}

The number of dropped packets is determined by the distribution of channel gain, which depends on the propagation environments and $N_{\rm t}$ as well. In Fig. \ref{fig:model}, we provide the number of dropped packets over Nakagami-$m$ fading channel with different values of $m$ and $N_{\rm t}$. We consider the worst case that all the users are located at the edge of the cell (i.e., user-BS distance is $200$~m). Since the average channel gains of all the users are the same, the total bandwidth and transmit power are equally allocated to all the users. Then, $N^{\rm c}_k = N^{\rm c}_{\max}/K$ and $P_k^{\rm th} = P^{\max}/K$. We set ${\varepsilon^c}={\varepsilon^q}={\varepsilon_{\rm D}}/3$. ${\varepsilon_k^h}$ is calculated from \eqref{eq:consteh}, where ${f_{\rm g}}\left( g \right) = \frac{{{{\left( {mg} \right)}^{m - 1}}}}{{\left( {m - 1} \right)!}}\exp \left( { - mg} \right)$ when $N_{\rm t} =1$ and $m > 1$\cite{WirelessCom} and $f_{\rm g}\left( g \right) = \frac{1}{{\left( {{N_\mathrm{t}} - 1} \right)!}}{g^{{N_\mathrm{t}} - 1}}{e^{ - g}}$ when $N_{\rm t} > 1$ and $m = 1$ \cite{Telatar1995Capacity}. All the results in the figure are obtained under constraint ${\varepsilon_k^h}\leq {\varepsilon_{\rm D}}/3$. The results when $N_{\rm t} = 1$ and $m=1$ are not shown, because constraint ${\varepsilon_k^h}\leq {\varepsilon_{\rm D}}/3$ cannot be satisfied under the transmit power constraint. The number of dropped packets in transmitting $10^{10}$ packets with proactive packet dropping policy is $10^{10}{\varepsilon_k^h}$. To show the performance gain of proactive packet dropping, we also provide the results for an intuitive packet dropping policy, which simply drops all the packets to the $k$th user when $g_k < \frac{{{N_0}B N^{\rm c}_k\gamma_k }}{{\mu_k {P_k^{\rm th }}}}$. We can see that proactive packet dropping policy can help reduce the number of dropped packets.

\section{Conclusions}
In this paper, we studied how to optimize resource allocation to guarantee ultra-low latency and ultra-high reliability for radio access networks in typical application scenarios where the required delay is shorter than channel coherence time. Both queuing delay and transmission delay were considered in the latency, and the transmission error probability, queueing delay violation probability, and packet dropping probability were taken into account in the reliability. We first showed that the required transmit power to ensure the QoS is unbounded when queueing delay bound is shorter than channel coherence time. To satisfy the QoS requirement with finite transmit power, a proactive packet dropping mechanism was proposed. A framework for optimizing resource allocation to ensure the stringent QoS was established, where a queue state and channel state information dependent transmit power allocation and packet dropping policies were optimized for single user case, and bandwidth allocation was further optimized for multi-user scenario,  to minimize the required maximal transmit power of the BS. How to apply the proposed framework to frequency-selective channel was also addressed. Simulation results validated that effective bandwidth can be used to optimize resource allocation for Poisson process, IPP and SPP, which are representative traffic models to characterizing performance of a system with queueing. Numerical results showed that the transmission error probability, queueing delay violation probability, and packet dropping probability are in the same order of magnitude, and setting the three packet loss probabilities equal will cause minor power loss.


\appendices
\section{Effective bandwidth of Several Relevant Arrival Processes}
\label{App:EB}
\renewcommand{\theequation}{A.\arabic{equation}}
\setcounter{equation}{0}
{\bf{Poisson arrival process:}} The effective bandwidth of Poisson process is given by
\begin{align}
E^{B}_k(\theta_k) = \frac{{{\lambda _k}}}{{T_{\rm f}{\theta _k}}}\left( {{e^{{\theta _k}}} - 1} \right)\; \text{(packets/s)}.\label{eq:EBPoisson}
\end{align}
Substituting \eqref{eq:EBPoisson} into \eqref{eq:delay}, we can obtain the required QoS exponent $\theta_k = \ln \left[\frac{T_{\rm f}\ln(1/\varepsilon_k^q)}{\lambda_k D^q_{\max}}+1 \right]$. Then, \eqref{eq:EBPoisson}  can be re-expressed as a function of $(D^q_{\max}, {\varepsilon_k^q})$ as
\begin{align}
E^{B}_k(\theta_k) = \frac{\ln(1/\varepsilon_k^q)}{D^q_{\max} \ln \left[\frac{T_{\rm f}\ln(1/\varepsilon_k^q)}{\lambda_k D^q_{\max}}+1\right]}\; \text{(packets/s)}.\label{eq:EBDepson}
\end{align}

{\bf{IPP:}} The effective bandwidth of the IPP can be expressed as \cite{Mustafa2016EB}
\begin{align}
E_k^B\left( {{\theta _k}} \right) = \frac{\Omega}{{2\theta_k T_{\rm f}}}\;
\text{(packet/s)},\label{eq:EBIPP}
\end{align}
where $\Omega \triangleq [\left( {{e^{\theta_k} } - 1} \right)\lambda _k^{{\rm{on}}} - \left( {\alpha  + \beta } \right) ]+\sqrt {{{\left[ {\left( {{e^{\theta_k} } - 1} \right)\lambda _k^{{\rm{on}}} - \left( {\alpha  + \beta } \right)} \right]}^2} + 4\alpha \left( {{e^{\theta_k} } - 1} \right)\lambda _k^{{\rm{on}}}}$. Substituting \eqref{eq:EBIPP} into \eqref{eq:delay}, the QoS exponent $\theta_k$ can be obtained from
$\Omega = \frac{-2T_{\rm f}\ln{\varepsilon_k^q }}{D^q_{\max}}$ numerically.

%

{\bf{SPP:}} Deriving the effective bandwidth of autocorrelated processes is much harder than that of renewal processes. To overcome this difficulty, we provide an upper bound of the effective bandwidth of SPP. Without loss of generality, we assume $\lambda^{\rm I}_k \leq \lambda^{\rm II}_k$.

Consider a Poisson process with average arrival rate $\lambda^{\rm II}_k$, the arrival rate in the first state of SPP is less than that of the Poisson process. Thus, the effective bandwidth of the SPP is less than that of the Poisson process, which can be obtained by substituting $\lambda_k = \lambda^{\rm II}_k$ into \eqref{eq:EBPoisson}.

\section{Upper bound of the packet dropping probability}
\label{App:UB}
\renewcommand{\theequation}{B.\arabic{equation}}
\setcounter{equation}{0}
\begin{proof}
To derive $\varepsilon_k^h$, we introduce an upper bound of $b_k^d\left( n \right)$ as follows,
\begin{align}
b_k^{U}\left( n \right) = \left\{ {\begin{array}{*{20}{c}}
{\max \left( {{T_\mathrm{f}}E_k^B({\theta}_k) - s_k^{\rm th },0} \right),{\rm{if }}\; {Q}_k\left( n \right) > 0,}\\
{0,\quad\quad\quad\quad\quad\quad\quad\quad\quad\quad{\rm{ if}}\;{{ {Q}_k\left( n \right)=0,}}}
\end{array}} \right. \nonumber
\end{align}
considering that $b_k^{U}\left( n \right) = b_k^d(n)$ when ${Q}_k\left( n \right) \geq {T_\mathrm{f}}E_k^B({\theta}) $ or ${Q}_k\left( n \right) = 0$, and $b_k^{U}\left( n \right) > b_k^d(n)$ when $0 < {Q}_k\left( n \right) < {T_\mathrm{f}}E_k^B({\theta}_k)$. Then, we can derive an upper bound of ${\mathbb{E}}[b_k^d\left( n \right) ]$ as
\begin{align}
{{\mathbb{E}}[b_k^U( n) ]} = \eta_k {\int_0^{\frac{{{N_0}{B N^{\rm c}_k}\gamma_k}}{{{\mu_k}P_k^{\rm th }}}} {( {T_\mathrm{f} E_k^B(\theta_k) - s_k^{\rm th }} ){f_{\rm g}}( g )dg} }.\nonumber
\end{align}
Substituting ${{\mathbb{E}}[b_k^U( n) ]}$ into \eqref{eq:defineh}, we obtain an upper bound of the packet dropping probability as
\begin{align}
\varepsilon_k^h \leq \int_0^{\frac{{{N_0}{B N^{\rm c}_k}\gamma_k}}{{{\mu_k}P_k^{\rm th }}}} {\left[ {1 - \frac{{s_k^{\rm th }}}{{{T_\mathrm{f}}E_k^B({\theta}_k)}}} \right]{f_{\rm g}}\left( g \right)dg} \label{eq:eh},
\end{align}
where $\eta_k = \Pr\{Q_k(n)>0\} = {{{\mathbb{E}}\{\sum\limits_{i \in {\mathcal{A}}_k} {{a_i}\left( n \right)}\}}}/{{\mathbb{E}}[s_k(n)]}= {{{\mathbb{E}}\{\sum\limits_{i \in {\mathcal{A}}_k} {{a_i}\left( n \right)}\}}}/{[{T_\mathrm{f}}E_k^B({\theta}_k)]}$ is applied.

By substituting $s_k^{\rm th }$ in \eqref{eq:sth} and considering \eqref{eq:power}, we have
\begin{align}
\frac{{{s_k^{\rm th }}}}{{{T_\mathrm{f}}{E_k^B}\left( \theta_k  \right)}}\approx \frac{\ln\left(1+\frac{\mu_k P_k^{\rm th} g_k}{N_0 B N^{\rm c}_k}\right)-\sqrt{\frac{V_k}{\phi B N^{\rm c}_k }}f^{-1}_{\rm Q}{(\varepsilon_k^c)}}{\ln\left(1+\gamma_k\right)-\sqrt{\frac{V_k}{\phi B N^{\rm c}_k }}f^{-1}_{\rm Q}{(\varepsilon_k^c)}} \label{eq:ratio}.
\end{align}
Because a packet is dropped only if it will be transmitted in deep fading, i.e. $g_k \to 0$, $V_k$ in \eqref{eq:dispersion} approaches $0$, and then \eqref{eq:ratio} can be further accurately approximated by
\begin{align}
\frac{{{s_k^{\rm th}}}}{{{T_\mathrm{f}}{E_k^B}\left( \theta_k\right)}} \approx \frac{\ln\left(1+\frac{\mu_k P^{\rm th}_k g_k}{N_0 B N^{\rm c}_k}\right)}{\ln\left(1+\gamma_k\right)}\label{eq:appratio}.
\end{align}

Substituting \eqref{eq:appratio} into \eqref{eq:eh}, we obtain the approximation in \eqref{eq:consteh}.
\end{proof}

\section{Proof of the convexity of the objective function in \eqref{eq:combine2}}
\label{App:combine}
\renewcommand{\theequation}{C.\arabic{equation}}
\setcounter{equation}{0}
\begin{proof}
For the Q-function  ${f_{\rm Q}}\left( x \right) = \frac{1}{{\sqrt {2\pi } }}\int_x^\infty  {\exp \left( { - \frac{{{\tau ^2}}}{2}} \right)} d\tau$, we have ${f'_{\rm Q}}\left( x \right) \buildrel \Delta \over =  - \frac{1}{{\sqrt {2\pi } }}{e^{ - {x^2}/2}} < 0$, and ${f''_{\rm Q}}\left( x \right) = \frac{x}{{\sqrt {2\pi } }}{e^{ - {x^2}/2}}>0$ when $x > 0$. Thus, ${f_{\rm Q}}\left( x \right)$ is an decreasing and strictly convex function when $x > 0$, i.e. ${f_{\rm Q}}\left( x \right) < 0.5$. Since the inverse function of a decreasing and strictly convex function is also strictly convex \cite{boyd}, $f_{\rm Q}^{ - 1}\left( {\varepsilon^c} \right)$ is strictly convex when $\varepsilon^c < 0.5$ (which is true for any application). Hence, the second term of \eqref{eq:combine2} is strictly convex.

To prove that the first term of \eqref{eq:combine2} is strictly convex, we first derive its second order derivative. Denote $y =  - \ln \left( {\varepsilon^q} \right)$ and $z = \frac{T_{\rm f}}{{D_{\max }^q{\lambda}}} > 0$. After removing the non-relevant constants, the first term of \eqref{eq:combine2} can be expressed as $f\left( y \right) = \frac{{ y}}{{\ln \left( {1 + z y} \right)}}$, and its
second order derivative is derived as
\begin{align}
\frac{{{d^2}f}}{{d{{\left( {\varepsilon^q} \right)}^2}}} = \left( {\frac{{{d^2}f}}{{d{y^2}}}} \right){\left( {\frac{{dy}}{{d\varepsilon^q}}} \right)^2} + \left( {\frac{{df}}{{dy}}} \right)\left( {\frac{{{d^2}y}}{{d{{\left( {\varepsilon^q} \right)}^2}}}} \right)\label{eq:secondorder}.
\end{align}
After some regular derivations, we can obtain that
\begin{align}
\frac{{dy}}{{d\varepsilon ^q}} =  - \frac{1}{{\varepsilon^q}}, \; \frac{{{d^2}y}}{{d{{\left( {\varepsilon^q} \right)}^2}}} = {\left( {\frac{1}{{\varepsilon^q}}} \right)^2}
,\label{eq:dfdy}
\end{align}
\begin{align}
&\frac{{df}}{{dy}} = \frac{{\left( {1 + zy} \right)\ln \left( {1 + zy} \right) - zy}}{{{{\left[ {\ln \left( {1 + zy} \right)} \right]}^2}\left( {1 + zy} \right)}}, \label{eq:dfdy2a}\\
&\frac{{{d^2}f}}{{d{y^2}}} = \frac{{2{z^2}y - \left( {2z + {z^2}y} \right)\ln \left( {1 + zy} \right)}}{{{{\left[ {\ln \left( {1 + zy} \right)} \right]}^3}{{\left( {1 + zy} \right)}^2}}}\label{eq:dfdy2b}.
\end{align}
After substituting \eqref{eq:dfdy}, \eqref{eq:dfdy2a} and \eqref{eq:dfdy2b} into \eqref{eq:secondorder}, we can finally obtain that
\begin{align}
\frac{{{d^2}f}}{{d{{\left( {\varepsilon^q} \right)}^2}}} &= \Big\{{{\left( {1 + zy} \right)}^2}{{\left[ {\ln \left( {1 + zy} \right)} \right]}^2} \nonumber\\
&- \left( {2z + zy + {z^2}y + {z^2}{y^2}} \right)\ln \left( {1 + zy} \right)+ 2{z^2}y\Big\}\times \nonumber\\
&\left\{{{{\left[ {\ln \left( {1 + zy} \right)} \right]}^3}{{\left( {1 + zy} \right)}^2}{{\left( {\varepsilon^q} \right)}^2}}\right\}^{-1}.\label{eq:finalresult}
\end{align}
Since the denominator is positive, we only need to show the numerator is positive. Denote the numerator of \eqref{eq:finalresult} as $f_{\rm mun}(x,z)$, where $x = yz$. Then, we have
\begin{align}
f_{\rm mun}(x,z) = &(1+x)^2\left[\ln(1+x)\right]^2 - (x+x^2)\ln(1+x) -\nonumber\\
&\left[(2+x)\ln(1+x) - 2x\right]z.\label{eq:numerator}
\end{align}
For $\varepsilon^q < 10^{-5}$, which is true for applications with ultra-high reliability requirement, $y > - \ln \left( 10^{-5}\right) > 10$, and then $x > 10 z$. Moreover, $(2+x)\ln(1+x) - 2x > 0, \forall x >0$. Then, we can obtain a lower bound of $f_{\rm mun}(x,z)$ as follows,
\begin{align}
f_{\rm LB}(x) = &(1+x)^2\left[\ln(1+x)\right]^2 - (x+x^2)\ln(1+x) - \nonumber\\
&\left[(2+x)\ln(1+x) - 2x\right]x/10.\label{eq:numeratorLB}
\end{align}
When $x = 0$, $f_{\rm LB}(x) = 0$. To prove $f_{\rm LB}(x) > 0, \forall x > 0$, we substitute $\nu = x + 1$ into \eqref{eq:numeratorLB} and prove $f'_{\rm LB}(\nu) > 0, \forall \nu > 1$. It is not hard to derive that
\begin{align}
f'_{\rm LB}(\nu) = \frac{20{\nu}^2(\ln\nu)^2 + (10\nu - 2{\nu}^2)\ln\nu + (3\nu-11)(\nu-1)}{10\nu}.\label{eq:numeratorLBd}
\end{align}
Denote the numerator of \eqref{eq:numeratorLBd} as $f_{\rm LBnum}(\nu)$, which equals zero when $\nu = 0$. Besides,
\begin{align}
f'_{\rm LBnum}(\nu) = 40 \nu (\ln\nu)^2 + (10+36 \nu)\ln\nu + 4(\nu-1) > 0,& \nonumber\\
\forall \nu > 1.&\nonumber
\end{align}
As a result, $f'_{\rm LB}(\nu) > 0$, and hence $f_{\rm LB}(x)$ increases with $x$. Therefore, we have $f_{\rm LB}(x)>0, \forall x > 0$. This completes the proof.
\end{proof}

\section{Proof of the optimality of the two-step method}
\label{App:optimal}
\renewcommand{\theequation}{D.\arabic{equation}}
\setcounter{equation}{0}
\begin{proof}
Denote an arbitrary feasible solution of problem \eqref{eq:epsilon3} and the related transmit power as $(\tilde{\varepsilon}^{q},\tilde{\varepsilon}^{c},\tilde{\varepsilon}^{h})$ and $\tilde{P}^{{\max}}$, respectively. Given $\tilde{\varepsilon}^{h}$, we can obtain the global minimal transmit power $P^{\max}(\tilde{\varepsilon}^{h}) \leq \tilde{P}^{{\max}}$ by solving problem \eqref{eq:combine2}, which is for Poisson arrival process. In the second step, the global optimal $\varepsilon^{h^*}$ is obtained such that $P^{\max^*} \leq P^{\max}(\tilde{\varepsilon}^{h})$. Therefore, $P^{\max^*} \leq \tilde{P}^{{\max}}$.
\end{proof}

\section{Achievable rate over frequency-selective channel}
\label{App:Nschannel}
\renewcommand{\theequation}{E.\arabic{equation}}
\setcounter{equation}{0}
Denote the channel vector on the $j$th subchannel of the $k$th  user as ${\bf{h}}_{kj} \in {\mathbb{C}}^{N_{\rm t} \times 1}$. Then, the channel matrix over frequency-selective channel is equivalent to a $N_{\rm t} N^{\rm sc}_k \times N^{\rm sc}_k $ MIMO channel with bandwidth $W_{\rm c}$, i.e., ${\bf{H}}_k = \text{diag}\left( {{{\bf{h}}_{k1}},{{\bf{h}}_{k2}},...,{{\bf{h}}_{k{N^{\rm sc}_k}}}} \right)$
and ${\bf{H}}_k^H{{\bf{H}}_k} = \text{diag}\left( {{g_{k1}},{g_{k2}},...,{g_{k{N^{\rm sc}_k}}}} \right)$, where $g_{kj} = {\bf{h}}^H_{kj}{\bf{h}}_{kj}$ is the channel gain on the $j$th subchannel allocated to the $k$th user and also one of the eigenvalues of ${\bf{H}}_k^H{{\bf{H}}_k} $. Then, by substituting the eigenvalues into (96) and  (97) in \cite{Yury2014Quasi}, the number of packets that can be transmitted in one frame can be expressed as \eqref{eq:Nkst}.

\bibliographystyle{IEEEtran}
\bibliography{ref}

\begin{thebibliography}{10}
\providecommand{\url}[1]{#1}
\csname url@samestyle\endcsname
\providecommand{\newblock}{\relax}
\providecommand{\bibinfo}[2]{#2}
\providecommand{\BIBentrySTDinterwordspacing}{\spaceskip=0pt\relax}
\providecommand{\BIBentryALTinterwordstretchfactor}{4}
\providecommand{\BIBentryALTinterwordspacing}{\spaceskip=\fontdimen2\font plus
\BIBentryALTinterwordstretchfactor\fontdimen3\font minus
  \fontdimen4\font\relax}
\providecommand{\BIBforeignlanguage}[2]{{%
\expandafter\ifx\csname l@#1\endcsname\relax
\typeout{** WARNING: IEEEtran.bst: No hyphenation pattern has been}%
\typeout{** loaded for the language `#1'. Using the pattern for}%
\typeout{** the default language instead.}%
\else
\language=\csname l@#1\endcsname
\fi
#2}}
\providecommand{\BIBdecl}{\relax}
\BIBdecl

\bibitem{CYGC16}
C.~She, C.~Yang, and T.~Quek, ``Cross-layer transmission design for tactile
  internet,'' in \emph{Proc. IEEE Globecom}, 2016.

\bibitem{3GPP2016Scenarios}
3GPP, \emph{Study on Scenarios and Requirements for Next Generation Access
  Technologies}.\hskip 1em plus 0.5em minus 0.4em\relax Technical Specification
  Group Radio Access Network, Technical Report 38.913, Release 14, Oct. 2016.

\bibitem{Gerhard2014The}
G.~P. Fettweis, ``The tactile internet: Applications \& challenges,''
  \emph{IEEE Vehic. Tech. Mag.}, vol.~9, no.~1, pp. 64--70, Mar. 2014.

\bibitem{Popovski2014METIS}
{P. Popovski, \emph{et al.}}, ``Deliverable d6.3 intermediate system evaluation
  results.''\hskip 1em plus 0.5em minus 0.4em\relax ICT-317669-METIS/D6.3,
  2014.

\bibitem{3GPPQoS}
3GPP, \emph{Further Advancements for E-{UTRA} Physical Layer Aspects}.\hskip
  1em plus 0.5em minus 0.4em\relax Technical Specification Group Radio Access
  Network, Technical Report 36.814, Release 9, Mar. 2010.

\bibitem{A2014Scenarios}
{A. Osseiran, F. Boccardi and V. Braun, \emph{et al.}}, ``Scenarios for 5{G}
  mobile and wireless communications: The vision of the {METIS} project,''
  \emph{IEEE Commun. Mag}, vol.~52, no.~5, pp. 26--35, May. 2014.

\bibitem{Shao2015ultra}
S.-Y. Lien, S.-C. Hung, K.-C. Chen, and Y.-C. Liang, ``Ultra-low-latency
  ubiquitous connections in heterogeneous cloud radio access networks,''
  \emph{IEEE Wireless Commun.}, vol.~22, no.~3, pp. 22--31, Jun. 2015.

\bibitem{Capozzi2013Downlink}
F.~Capozzi, G.~Piro, L.~Grieco, G.~Boggia, and P.~Camarda, ``Downlink packet
  scheduling in {LTE} cellular networks: Key design issues and a survey,''
  \emph{IEEE Commun. Surveys Tuts.}, vol.~15, no.~2, pp. 678--700, 2013.

\bibitem{Meryem2016Tactile}
M.~Simsek, A.~Aijaz, M.~Dohler, J.~Sachs, and G.~Fettweis, ``5{G}-enabled
  tactile internet,'' \emph{IEEE J. Select. Areas Commun.}, vol.~34, no.~3, pp.
  460--473, Mar. 2016.

\bibitem{Shehzad2015Control}
S.~A. Ashraf, F.~Lindqvist, R.~Baldemair, and B.~Lindoff, ``Control channel
  design trade-offs for ultra-reliable and low-latency communication system,''
  in \emph{IEEE Globecom Workshops}, 2015.

\bibitem{Petteri2015A}
{P. Kela and J. Turkka, \emph{et al.}}, ``A novel radio frame structure for
  5{G} dense outdoor radio access networks,'' in \emph{Proc. IEEE VTC Spring},
  2015.

\bibitem{Yury2010Channel}
Y.~Polyanskiy, H.~V. Poor, and S.~Verd\'{u}, ``Channel coding rate in the
  finite blocklength regime,'' \emph{IEEE Trans. Inf. Theory}, vol.~56, no.~5,
  pp. 2307--2359, May 2010.

\bibitem{Kai2014Polar}
K.~Niu, K.~Chen, J.~Lin, and Q.~T. Zhang, ``Polar codes: Primary concepts and
  practical decoding algorithms,'' \emph{IEEE Commun. Mag}, vol.~52, no.~7, pp.
  192--203, Jul. 2014.

\bibitem{David2014Achieving}
D.~Ohmann, M.~Simsek, and G.~P. Fettweis, ``Achieving high availability in
  wireless networks by an optimal number of {R}ayleigh-fading links,'' in
  \emph{IEEE Globecom Workshops}, 2014.

\bibitem{Felix2015diversity}
F.~Kirsten, D.~Ohmann, M.~Simsek, and G.~P. Fettweis, ``On the utility of
  macro- and microdiversity for achieving high availability in wireless
  networks,'' in \emph{Proc. IEEE PIMRC}, 2015.

\bibitem{Beatriz2015Reliable}
G.~Pocovi, B.~Soret, M.~Lauridsen, K.~I. Pedersen, and P.~Mogensen, ``Signal
  quality outage analysis for ultra-reliable communications in cellular
  networks,'' in \emph{IEEE Globecom Workshops}, 2015.

\bibitem{Osman2015Analysis}
O.~N.~C. Yilmaz, Y.-P.~E. Wang, N.~A. Johansson, N.~Brahmi, S.~A. Ashraf, and
  J.~Sachs, ``Analysis of ultra-reliable and low-latency 5{G} communication for
  a factory automation use case,'' in \emph{IEEE ICC Workshops}, 2015.

\bibitem{Niklas2015Radio}
N.~A. Johansson, Y.-P.~E. Wang, E.~Eriksson, and M.~Hessler, ``Radio access for
  ultra-reliable and low-latency 5{G} communications,'' in \emph{IEEE ICC
  Workshops}, 2015.

\bibitem{EB}
C.~Chang and J.~A. Thomas, ``Effective bandwidth in high-speed digital
  networks,'' \emph{IEEE J. Sel. Areas Commun.}, vol.~13, no.~6, pp.
  1091--1100, Aug. 1995.

\bibitem{EC}
D.~Wu and R.~Negi, ``Effective capacity: A wireless link model for support of
  quality of service,'' \emph{IEEE Trans. Wireless Commun.}, vol.~2, no.~4, pp.
  630--643, Jul. 2003.

\bibitem{Beatriz2014Tradeoffs}
B.~Soret, P.~Mogensen, K.~I. Pedersen, and M.~C. Aguayo-Torres, ``Fundamental
  tradeoffs among reliability, latency and throughput in cellular networks,''
  in \emph{IEEE Globecom Workshops}, Dec. 2014.

\bibitem{Adnan2016Towards}
A.~Aijaz, ``Towards 5{G}-enabled tactile internet: Radio resource allocation
  for haptic communications,'' in \emph{Proc. IEEE WCNC}, 2016.

\bibitem{Gross2015Delay}
S.~Schiessl, J.~Gross, and H.~Al-Zubaidy, ``Delay analysis for wireless fading
  channels with finite blocklength channel coding,'' in \emph{Proc. ACM MSWiM},
  2015.

\bibitem{Throughput2011Mustafa}
M.~C. Gursoy, ``Throughput analysis of buffer-constrained wireless systems in
  the finite blocklength regime,'' in \emph{Proc. IEEE ICC}, 2011.

\bibitem{Shengfeng2015Convexity}
S.~Xu, T.-H. Chang, S.-C. Lin, C.~Shen, and G.~Zhu, ``On the convexity of
  energy-efficient packet scheduling problem with finite blocklength codes,''
  in \emph{IEEE Globecom Workshops}, 2015.

\bibitem{Berry2013}
R.~A. Berry, ``Optimal power-delay tradeoffs in fading channels---small-delay
  asymptotics,'' \emph{IEEE Trans. Inf. Theory}, vol.~59, no.~6, pp.
  3939--3952, Jun. 2013.

\bibitem{Yury2014Quasi}
W.~Yang, G.~Durisi, T.~Koch, and Y.~Polyanskiy, ``Quasi-static multiple-antenna
  fading channels at finite blocklength,'' \emph{IEEE Trans. Inf. Theory},
  vol.~60, no.~7, pp. 4232--4264, Jul. 2014.

\bibitem{Ward1993Tail}
W.~Whitt, ``Tail probabilities with statistical multiplexing and effective
  bandwidths in multi-class queues,'' \emph{Telecommunication Systems}, vol.~2,
  no.~1, pp. 71--107, 1993.

\bibitem{Jian2015IPP}
J.~Wu, Y.~Bao, G.~Miao, S.~Zhou, and Z.~Niu, ``Base station sleeping control
  and power matching for energy-delay tradeoffs with bursty traffic,''
  \emph{IEEE Trans. Veh. Technol.}, vol.~65, no.~5, pp. 3657--3675, May 2016.

\bibitem{Tony2016Backhaul}
G.~Zhang, T.~Q.~S. Quek, M.~Kountouris, A.~Huang, and H.~Shan, ``Fundamentals
  of heterogeneous backhaul design---analysis and optimization,'' \emph{IEEE
  Trans. Commun.}, vol.~64, no.~2, pp. 876--889, Feb. 2016.

\bibitem{Daquan2014Device}
D.~Feng, L.~Lu, Y.~Yuan-Wu, G.~Y. Li, S.~Li, and G.~Feng, ``Device-to-device
  communications in cellular networks,'' \emph{IEEE Commun. Mag.}, vol.~52,
  no.~4, pp. 49--55, Apr. 2014.

\bibitem{She2016GCworkshop}
C.~She, C.~Yang, and T.~Q.~S. Quek, ``Uplink transmission design with massive
  machine type devices in tactile internet,'' in \emph{IEEE Globecom
  Workshops}, 2016.

\bibitem{Tang2007Quality}
J.~Tang and X.~Zhang, ``Quality-of-service driven power and rate adaptation
  over wireless links,'' \emph{IEEE Trans. Wireless Commun.}, vol.~6, no.~8,
  pp. 3058--3068, Aug. 2007.

\bibitem{Mehdi2013Performance}
M.~Khabazian, S.~Aissa, and M.~Mehmet-Ali, ``Performance modeling of safety
  messages broadcast in vehicular ad hoc networks,'' \emph{IEEE Trans. Intell.
  Transp. Syst.}, vol.~14, no.~1, pp. 380--387, Mar. 2013.

\bibitem{3GPP2012MTC}
G.~R1-120056, ``Analysis on traffic model and characteristics for {MTC} and
  text proposal.''\hskip 1em plus 0.5em minus 0.4em\relax Technical Report,
  TSG-RAN Meeting WG1\#68, Dresden, Germany, 2012.

\bibitem{Gross1985MD1}
D.~Gross and C.~Harris, \emph{Fundamentals of Queueing Theory}.\hskip 1em plus
  0.5em minus 0.4em\relax Wiley, 1985.

\bibitem{Hassan2013A}
H.~A. Omar, W.~Zhuang, A.~Abdrabou, and L.~Li, ``Performance evaluation of
  {V}e{MAC} supporting safety applications in vehicular networks,'' \emph{IEEE
  Trans. Emerg. Topics Comput.}, vol.~1, no.~1, pp. 69--83, Aug. 2013.

\bibitem{squeezing1996}
G.~L. Choudhury, D.~M. Lucantoni, and W.~Whitt, ``Squeezing the most out of
  {ATM},'' \emph{IEEE Trans. Commun.}, vol.~44, no.~2, pp. 203--217, Feb. 1996.

\bibitem{Telatar1995Capacity}
I.~E. Telatar, \emph{Capacity of multi-antenna Gaussian channels}, 1995.

\bibitem{WirelessCom}
A.~Goldsmith, \emph{Wireless Communications}.\hskip 1em plus 0.5em minus
  0.4em\relax Cambridge University Press, 2005.

\bibitem{Kumar2009QoS}
I.-H. Hou, V.~Borkar, and P.~R. Kumar, ``A theory of {Q}o{S} for wireless,'' in
  \emph{Proc. IEEE INFOCOM}, 2009.

\bibitem{boyd}
S.~Boyd and L.~Vandanberghe, \emph{{C}onvex {O}ptimization}.\hskip 1em plus
  0.5em minus 0.4em\relax Cambridge Univ. Press, 2004.

\bibitem{Giuseppe2016Toward}
G.~Durisi, T.~Koch, and P.~Popovski, ``Toward massive, ultrareliable, and
  low-latency wireless communication with short packets,'' \emph{Proc. IEEE},
  vol. 104, no.~9, pp. 1711--1726, Aug. 2016.

\bibitem{ECMIMO}
B.~Soret, M.~C. Aguayo-Torres, and J.~T. Entrambasaguas, ``Capacity with
  explicit delay guarantees for generic sources over correlated {R}ayleigh
  channel,'' \emph{IEEE Trans. Wireless Commun.}, vol.~9, no.~6, pp.
  1901--1911, Jun. 2010.

\bibitem{Mustafa2016EB}
M.~Ozmen and M.~C. Gursoy, ``Wireless throughput and energy efficiency with
  random arrivals and statistical queuing constraints,'' \emph{IEEE Trans. Inf.
  Theory}, vol.~62, no.~3, pp. 1375--1395, Mar. 2016.

\end{thebibliography}

\end{document}